\documentstyle[12pt]{article}
\newcommand{\newsection}[1]{
 \vspace{10mm} \pagebreak[3]
 \addtocounter{section}{1}
 \setcounter{subsection}{0}
 \setcounter{paragraph}{0}
 \setcounter{equation}{0}
 \setcounter{figure}{0}
 \setcounter{table}{0}
 \addcontentsline{toc}{section}{\protect\numberline{\arabic{section}}{#1}}
 \begin{flushleft}
  {\large\bf \thesection. #1}
 \end{flushleft}
 \nopagebreak}

\def\al{\alpha}
\def\bt{\beta}
\def\gm{\gamma}                \def\Gm{\Gamma}
\def\dl{\delta}                
\def\ep{\epsilon}

\def\lm{\lambda}               

\def\th{\theta}               
\def\vph{\varphi}

\def\om{\omega}               \def\Om{\Omega}
\def\sg{\sigma}               \def\Sg{\Sigma}

\def\Ac{\mbox{\protect$\cal A$}}
\def\Bc{\mbox{\protect$\cal B$}}
\def\Dc{\mbox{\protect$\cal D$}}
\def\Jc{\mbox{\protect$\cal J$}}
\def\Oc{\mbox{\protect$\cal O$}}
\def\Sc{\mbox{\protect$\cal S$}}

\def\Ab{{\bf A}}

\def\FF{I\!\!F}
\def\RR{I\!\!R}
\def\ZZ{Z\!\!\! Z}

\def\sp{{\rm sp}}
\def\ad{{\rm ad}}

\def\pt{\partial}
\def\goto{\rightarrow}
\def\del{\nabla}
\def\wh#1{\widehat{(#1)}}
\def\bup#1{{\hat{#1}}}  
\def\bdn#1{{\check{#1}}}  
\def\inv{^{-1}}
\def\sds{\subset\hskip -1em +}  
\def\sdp{\subset\hskip -1em \times} 

\def\half{{\scriptstyle {1 \over 2}}}

\def\hs{\hspace{5mm}}
\def\hsc{\hspace{5mm},\hspace{5mm}}
\def\ie{{\em i.e.\ }}
\def\eg{{\em e.g.\ }}

\def\beq{\begin{equation}}
\def\eeq{\end{equation}}

\begin{document}


\begin{titlepage}

\begin{flushright}
RI-8-95\\
hep-th/9509013\\[5mm]
September 1995
\end{flushright}

\begin{center}
\Large
WZNW Models and Gauged WZNW Models \\
Based On \\
a Family of Solvable Lie Algebras
\\[10mm]
\large
Amit Giveon\footnote{E-mail: giveon@vms.huji.ac.il },
Oskar Pelc\footnote{E-mail: oskar@shum.cc.huji.ac.il}
\normalsize and \large
Eliezer Rabinovici\footnote{E-mail: eliezer@vms.huji.ac.il}
\normalsize
\\[5mm]
{\em Racah Institute of Physics, The Hebrew University\\
  Jerusalem, 91904, Israel}
\\[15mm]
\end{center}

\begin{abstract}
A family of solvable self-dual Lie algebras that are not double extensions
of Abelian algebras and, therefore, cannot be obtained through a Wigner
contraction, is presented. We construct WZNW and gauged WZNW
models based on the first two algebras in this family. We also analyze
some general phenomena arising in such models.
\end{abstract}

\end{titlepage}

\flushbottom


\newsection{Introduction}

WZNW models \cite{WZNW} and gauged WZNW models \cite{GWZNW} have served as
building blocks of various string theories.
For the construction of a WZNW model, one
needs a Lie group, and also a {\em metric} -- a non-degenerate, symmetric
bilinear form -- on the corresponding Lie algebra. A Lie algebra that admits
such a metric is called {\em self-dual}.

At first {\em reductive} algebras (direct sums
of semi-simple and Abelian algebras) were considered. Such algebras have
natural candidates for the invariant metric -- the Killing form for the
semi-simple part and an arbitrary metric for the Abelian part. However,
there exist self-dual algebras that are not reductive and these also can be
used for the construction of WZNW models (and their supersymmetric extensions)
\cite{Nappi-Witten}%
-\cite{FOF}.

One of the interesting features of an affine Sugawara construction based on a
non-semi-simple (and {\em indecomposable} -- not an orthogonal direst sum)
algebra is that the resulting central charge is integer and equals to the
dimension of the algebra \cite{ORS}\cite{FOF-Stan}. This may be a sign of some
interesting phenomena.
When the algebra can be obtained through a Wigner contraction \cite{Ino-Wig}
of a semi-simple algebra, as described in \cite{ORS},
this is explained by the fact that in the contraction process, a
semi-classical limit is taken -- the levels of the simple components are
taken to infinity. This suggests that the resulting model has a free-field
representation (this was demonstrated for the example of \cite{Nappi-Witten}
in \cite{Kiritsis-Kounnas}).

In this paper we consider more examples of non-reductive self-dual algebras
and study the $\sg$-model string backgrounds that correspond to them.

Any non-reductive (indecomposable) self-dual algebra can be constructed,
starting from an Abelian algebra, by a sequence of construction steps, each
of which is either an (orthogonal) direct product or a procedure called
``double extension'' \cite{Med-Rev}. If such an algebra can be obtained
through a Wigner contraction, it is necessarily a double extension of an
Abelian algebra, \ie, it is a result of a {\em single-step sequence}.
All the algebras used so far to construct WZNW models are double
extensions of Abelian algebras and, therefore, possibly can be obtained
through a Wigner contraction (for some of them this was explicitly shown
\cite{ORS}\cite{FOF-Stan}).
One might have suspected that {\em all} non-reductive, self-dual algebras
can be obtained through a Wigner contraction. If this was true, it would have
a significant implication on the structure of such algebras and the
WZNW models based on them. It turns out, however, that this is not true.

In fact, we present an infinite family of indecomposable, non-reductive,
self-dual algebras $\{\Ac_{3m}\}$ \cite{OSP}, that (except $\Ac_3$ and
$\Ac_6$) are {\em not} double extensions of Abelian algebras and, therefore,
{\em cannot} be obtained through a Wigner contraction
($\Ac_6$ is also unobtainable through a Wigner contraction). The algebra
\[ \Ac_n\equiv \sp\{T_i\}_{0\leq i\leq n} \]
is defined by the Lie bracket
\beq\label{lie}
  [T_i,T_j]=\left\{\begin{array}{cc}
    \wh{i-j}T_{i+j}	&	i+j\leq n \\
    0			&	\mbox{otherwise}.
  \end{array}\right.
\eeq
where $\hat{i}\equiv i \bmod 3$ is chosen to be in $\{-1,0,1\}$.
When $\hat{n}=0$, the metric
\beq (T_i,T_j)=\dl_{i+j-n}+b\dl_i\dl_j \eeq
is an invariant metric on $\Ac_n$ (for arbitrary $b$).
We construct WZNW and gauged WZNW models based on the first two
algebras in this series: $\Ac_3$ and $\Ac_6$.

The paper is organized as follows: in section 2, we describe the family
$\{\Ac_n\}$ of self-dual algebras. While
constructing the gauged WZNW examples, we encountered some phenomena
which are very common when the algebras are not semi-simple and seldom
(or never) appear otherwise. In section 3, we analyze some of them in a
general setting: the appearance of constraints; {\em singular} gauging
(when the restriction of the metric to the gauged subalgebra is singular%
\footnote{For examples of the special case of {\em null} gauging, when the
  metric on the gauged subalgebra vanishes, see \cite{Null} and references
  therein.})
and the gauging of a central subgroup. In section 4, we construct WZNW and
gauged WZNW models based on $\Ac_3$ and $\Ac_6$.
In Appendix A, we collect some parenthetical
remarks complementing the main text and, in Appendix B, we list the
geometrical information of all the $\sg$ models derived in section 4.
For all these models the one-loop beta functions \cite{beta} vanish and
the central charge is equal to the dimension of the $\sg$-model target
manifold.

\newsection{A New Family of Solvable Self-Dual Lie Algebras}

In this section we describe the family $\{\Ac_n\}$ of self-dual algebras,
as obtained in \cite{OSP}. The main results are described in the
introduction and the reader interested in the physical results may skip
directly to section 3.
We start, in subsection 2.1 with a review of the two methods for
constructing self-dual Lie algebras -- a double extension and a Wigner
contraction. In subsection 2.2 we define the
algebras $\Ac_n$ and prove that (for $\hat{n}=0$) these are indeed
self-dual Lie algebras. In subsection 2.3 we find all the ideals of $\Ac_n$.
This result is used in the last subsection, where we check which of these
algebras is a double extension of an Abelian algebra or a result of a
Wigner contraction.

\subsection{The Construction of Self-Dual Lie algebras}

A {\em self-dual Lie algebra} $\Ac$ is a Lie algebra that admits an
{\em invariant metric}, \ie a symmetric non-degenerate bilinear form
$(\cdot,\cdot)$ which
is invariant under the adjoint action of the corresponding group:
\beq (gx_1g\inv,gx_2g\inv)=(x_1,x_2), \hs \forall x_i\in\Ac \eeq
for any $g$ in the group, or equivalently:
\beq\label{invariance}
  ([y,x_1],x_2)=-(x_1,[y,x_2]), \hs \forall x_i\in\Ac
\eeq
for any $y\in\Ac$. The best known families of self-dual algebras are
the semi-simple algebras (where the (unique) invariant metric is the
Killing form) and the Abelian algebras (for which every metric is
trivially invariant). However, these are not the only ones. In this
section we are concerned with the search for self dual algebras that are
neither semi-simple nor Abelian. Given two self dual algebras, their direct
sum equipped with the natural direct sum metric, is also self dual (this
construction will be called an {\em orthogonal direct sum}),
therefore, in the construction of self dual algebras, the non-trivial task
is to find the {\em indecomposable} ones, \ie algebras that are not
orthogonal direct sums.
It has been shown \cite{Med-Rev} that any indecomposable self-dual Lie
algebra, which is neither simple nor one dimensional, is a {\em double
extension} of a smaller self-dual Lie algebra (see also \cite{FOF-Stan}),
therefore, one may attempt to use the procedure of double extension
for actual construction of new indecomposable self-dual Lie algebras.

The {\em double extension of a self-dual Lie algebra $\Ac$ by another
Lie algebra $\Bc$} (not necessarily self-dual) can be seen as a two-step
process. The first step is to combine them to a {\em semi-direct sum}%
\footnote{By this we mean that the vector space $\Sc$ is a direct sum of
  the vector spaces $\Bc$ and $\Ac$, $\Bc$ is a subalgebra of $\Sc$:
  $[\Bc,\Bc]\subset\Bc$ and $\Ac$ is an ideal of $\Sc$:
  $[\Sc,\Ac]\subset\Ac$.}
\beq \Sc=\Bc\sds\Ac \eeq
in such a way that the metric in $\Ac$ will be invariant also under the
action of $\Bc$. For this, one needs, in addition to the algebras $\Ac$
and $\Bc$, an action (representation) of $\Bc$ in $\Ac$
\beq\label{yyx}
  y:x\goto[y,x],\hs [[y_1,y_2],x]=[y_1,[y_2,x]]-[y_2,[y_1,x]]
\eeq
that will satisfy the mixed Jacobi identities
\beq\label{yxx} [y,[x_1,x_2]]=[[y,x_1],x_2]+[x_1,[y,x_2]] \eeq
and the invariance condition (\ref{invariance}) (here $x,x_i\in\Ac$,
$y,y_a\in\Bc$). Given bases $\{x_i\}$ and
$\{y_a\}$ for $\Ac$ and $\Bc$ respectively, the Lie bracket of $\Sc$ is
of the form
\beq\label{LieS} \begin{array}{c|cc}
  \rule[-4mm]{0mm}{8mm}
  [\cdot,\cdot] &  y_b            & x_j          \\
  \hline\rule[-4mm]{0mm}{10mm}
            y_a &  {f_{ab}}^c y_c & {f_{aj}}^k x_k \\
  \rule[-4mm]{0mm}{8mm}
            x_i & -{f_{bi}}^k x_k & {f_{ij}}^k x_k
\end{array}\eeq
${f_{ij}}^k$ and ${f_{ab}}^c$ are the structure constants of $\Ac$ and
$\Bc$ respectively and as such satisfy the Jacobi Identities; ${f_{ij}}^k$
satisfies an additional identity
\beq {f_{ij}}^l\Om_{lk}+{f_{ik}}^l\Om_{lj}=0, \eeq
expressing the invariance (\ref{invariance}) of the metric
$\Om_{ij}=(x_i,x_j)$ on $\Ac$;
${f_{aj}}^k$ represent the action of $y_a$ in $\Ac$ and identities
(\ref{yyx}), (\ref{yxx}) and (\ref{invariance}) take respectively the form
\beq\label{yyx-coo}
  {f_{ab}}^c{f_{ck}}^l={f_{am}}^l{f_{bk}}^m-{f_{bm}}^l{f_{ak}}^m,
\eeq
\beq\label{yxx-coo}
  {f_{ak}}^l{f_{ij}}^k={f_{ai}}^k{f_{kj}}^l+{f_{ik}}^l{f_{aj}}^k
\eeq
and
\beq\label{inv-coo} {f_{aj}}^l\Om_{lk}+{f_{ak}}^l\Om_{lj}=0. \eeq

The second step is the extension of $\Sc$ by an Abelian algebra
$\Bc^*$ with $\dim\Bc^*=\dim\Bc$. This step is completely determined
by the first one (the Lie bracket in $\Sc$ and the metric on $\Ac$)
and in an appropriate basis $\{z^a\}$ for $\Bc^*$ the Lie bracket of
the resulting algebra, which will be denoted by $\Dc$, is
\beq\label{LieD} \begin{array}{c|ccc}
  \rule[-4mm]{0mm}{8mm}
  [\cdot,\cdot] &  y_b            & x_j            & z^b \\ \hline
  \rule[-4mm]{0mm}{10mm}
            y_a &  {f_{ab}}^c y_c & {f_{aj}}^k x_k & -{f_{ac}}^b z^c \\
  \rule[-4mm]{0mm}{8mm}
            x_i & -{f_{bi}}^k x_k & {f_{ij}}^kx_k+{f_{ci}}^k\Om_{kj} z^c
                                                   & 0 \\
  \rule[-4mm]{0mm}{8mm}
            z^a & {f_{bc}}^a z^c  & 0              & 0
\end{array}\eeq
This algebra has an invariant metric, which in the above basis is
\beq
  (\cdot,\cdot)=\left(\begin{array}{ccc}
     \om_{ab} & 0        & \dl_a^b \\
     0        & \Om_{ij} & 0 \\
     \dl^a_b  & 0             & 0
  \end{array}\right),
\eeq
where $\om_{ab}$ is some invariant symmetric bilinear form on $\Bc$
(possibly degenerate; \eg the Killing form or zero).

The theorem proved in \cite{Med-Rev} states that an indecomposable
self-dual Lie algebra, which is neither simple nor one dimensional,
is a double extension of a self-dual algebra (with smaller dimension)
by a simple or one-dimensional algebra. Although this is a very
important and useful result for a general study of these algebras,
its straightforward application to an actual construction is
cumbersome, as we now explain. To double-extend a self-dual Lie
algebra $\Ac$, one needs (a Lie algebra $\Bc$ of) linear transformations
in $\Ac$ satisfying (\ref{yxx}) and (\ref{invariance}). One could, of
course, take the trivial action: $y:\;x\goto0$, but the resulting
algebra $\Dc$ is decomposable -- the original algebra $\Ac$ factorizes
out:
\beq \Dc=\Ac\stackrel{\perp}{\oplus}(\Bc\sds\Bc^*). \eeq
Moreover, it was shown in \cite{FOF-Stan} that even if $\Bc$ acts
non-trivially but its action is through {\em inner} derivations (\ie
the action of each $y\in\Bc$ coincides with the adjoint action of an
element $\hat{y}\in\Ac$: $y:x\goto[\hat{y},x]$), the result is also
decomposable. This means that for the construction of an indecomposable
double extension, one needs knowledge about the {\em outer}
(non-inner) derivations in $\Ac$ and such information is not available
in general. In the absence of general results, the suitable
transformations must be found by direct calculation. Given $\Ac$ and
$(\cdot,\cdot)_{\Ac}$, one must find solutions ${f_{aj}}^k$ of the
(linear) equations (\ref{yxx-coo}) and (\ref{inv-coo}) and identify
among them, by elimination, those that correspond to inner derivations.%

Another method for constructing new self-dual Lie algebras is by
performing a {\em Wigner contraction} \cite{Ino-Wig} (this was proposed, in
the context of WZNW models, in \cite{ORS}).
The initial data for this construction consists of a self dual Lie
algebra $\Sc_0$ and a sub algebra $\Bc_0$ of $\Sc_0$ such that the
restriction
of the metric $(\cdot,\cdot)$ on $\Sc_0$ to $\Bc_0$ is non-degenerate.
The last condition is equivalent to
\beq \Sc_0=\Bc_0\oplus\Bc_0^{\bot} \eeq
($\Bc_0^{\bot}$ is the orthogonal complement of $\Bc_0$ with respect
to the metric), therefore, bases $\{b_a^0\}$ for $\Bc_0$ and $\{a_i\}$
for $\Bc_0^{\bot}$ combine to a basis for $\Sc_0$. In this basis, the
Lie bracket in $\Sc_0$ has the general form%
\footnote{The structure constants satisfy the Jacobi identities and the
  identities expressing the invariance of the metric. In particular, the
  vanishing of ${f_{aj}}^c$ follows from the existence of a non-singular
  invariant metric:
  \[ (b_c^0,[b_a^0,a_j])=([b_c^0,b_a^0],a_j)=0. \]}
\beq\label{LieSa} [a_i,a_j]={f_{ij}}^k a_k+{f_{ij}}^c b_c^0, \eeq
\beq\label{LieSb}
  [b_a^0,a_j]={f_{aj}}^k a_k \hsc
  [b_a^0,b_b^0]={f_{ab}}^c b_c^0
\eeq
and the metric is
\beq
  (\cdot,\cdot)=
  \left(\begin{array}{cc}\Om_{ij}&0\\0&\Om_{ab}\end{array}\right).
\eeq

One now performs a Wigner contraction \cite{Ino-Wig} of
$\Sc_0\oplus\Bc_1$ (where $\Bc_1$ is isomorphic to $\Bc_0$ and
commutes with $\Sc_0$): define
\beq
  x_i\equiv\ep a_i, \hs
  y_a\equiv b_a^0+b_a^1, \hs
  z_a\equiv \half\ep^2(b_a^0-b_a^1),
\eeq
express the Lie brackets in terms of the new variables and take the
limit $\ep\goto0$. Since this is a singular limit, one obtains a {\em
new} algebra $\Dc$ (not isomorphic to $\Sc_0\oplus\Bc_1$) with the
following Lie brackets:
\beq\label{LieORS} \begin{array}{c|ccc}
  \rule[-4mm]{0mm}{8mm}
  [\cdot,\cdot] &  y_b          & x_j          & z_b \\ \hline
  \rule[-4mm]{0mm}{10mm}
            y_a &  {f_{ab}}^c y_c & {f_{aj}}^k x_k & {f_{ab}}^c z_c \\
  \rule[-4mm]{0mm}{8mm}
            x_i & -{f_{bi}}^k x_k & {f_{ij}}^c z_c & 0 \\
  \rule[-4mm]{0mm}{8mm}
            z_a & -{f_{ba}}^c z_c & 0            & 0
\end{array}\eeq
To obtain an invariant metric for $\Dc$ one starts with the natural
invariant metric on $\Sc_0\oplus\Bc_1$, that in the basis
$\{b_a^0,a_i,b_a^1\}$ is
\beq
  (\cdot,\cdot)_{\ep}=\left(\begin{array}{ccc}
     \bt_0\Om_{ab} & 0             & 0 \\
     0             & \bt_0\Om_{ij} & 0 \\
     0             & 0             & \bt_1\Om_{ab}
  \end{array}\right)
\eeq
(with arbitrary $\bt_0$, $\bt_1$). In the limit $\ep\goto0$ one obtains:
\beq\label{metricORS}
  (\cdot,\cdot)_0=\left(\begin{array}{ccc}
     \bt'\Om_{ab} & 0           & \bt\Om_{ab} \\
     0            & \bt\Om_{ij} & 0 \\
     \bt\Om_{ab}  & 0             & 0
  \end{array}\right)
\eeq
(in the basis $\{y_a,a_i,z_a\}$) where\footnote%
 {Note that to obtain a non-degenerate metric, $\bt_0$ and $\bt_1$
  must depend non-trivially on $\ep$. In fact they must diverge in the
  limit $\ep\goto0$.}
\beq
  \bt=\lim_{\ep\goto0}\half\ep^2(\bt_0-\bt_1) \hsc
  \bt'=\lim_{\ep\goto0}(\bt_0+\bt_1)
\eeq
and this form is invariant (by continuity) and, for $\bt\neq0$,
non-degenerate.

The resemblance of the resulting algebra to the one obtained by double
extension is apparent and indeed, using the metric to raise and lower
indexes, one immediately identifies it as the double extension of an
Abelian algebra $\Ac=\sp\{x_i\}$ by $\Bc=\sp\{y_a\}$ \cite{FOF-Stan}
(where $\sp\{x_i\}$ denotes the linear span of the set $\{x_i\}$).
However, this method has a clear advantage. Unlike double extension,
the initial data needed is very simple and generally available,
therefore, the method can be easily used to find many new non-trivial
self-dual algebras.

A natural question is if there are non-semi-simple,
indecomposable, self-dual algebras, that cannot be obtained by a Wigner
contraction. Any self-dual Lie algebra obtained through a Wigner
contraction can be obtained from an Abelian algebra by a {\em single}
double-extension, therefore, this question is closely related to the
problem of finding
(non-semi-simple, indecomposable, self-dual) algebras that their
construction out of simple and one dimensional algebras involves
more than one double extension%
\footnote{One might also consider a double
  extension of a {\em reductive} algebra $\Ac$ \ie an orthogonal direct
  sum of Abelian and semi-simple algebras. However, as shown in
  \cite{FOF-Stan}, the semi-simple factor of $\Ac$ factorizes also in
  the result $\Dc$ (because a semi-simple algebra does not have outer
  derivations), therefore, the result in this case is decomposable.},
and in this sense, are called {\em deeper} algebras \cite{FOF-Stan}.
As explained above, deeper algebras are not easy to find. In fact,
among the self dual algebras with dimension at most 5 (enumerated in
\cite{Kehagias}), none is deeper than a double extension of an
Abelian algebra. In the rest of this section we introduce and explore
a family of deeper algebras.

\subsection{The Algebra $\Ac_n$}
\label{An}

Consider a vector space, equipped with the basis $\{T_i\}_{i\in\ZZ}$
and the following ``Lie bracket'' \cite{OSP}:
\beq\label{Bracket-An} [T_i,T_j]=\wh{i-j}T_{i+j} \eeq
where $\hat{i}\equiv i \bmod 3$ is chosen to be in $\{-1,0,1\}$. The map
$i\goto\hat{i}$ is {\em almost} a ring homomorphism $\ZZ\goto\ZZ$: it
preserves multiplication
\beq\label{mult} \wh{ij}=\hat{i}\hat{j}, \eeq
and almost preserves addition%
\footnote{The $\Longleftarrow$ direction of (\ref{add2}) follows from
  (\ref{add1}), but for the other direction (\ref{add1}) only implies
  \[ \wh{i-j}=0 \Longrightarrow \hat{i}-\hat{j}=0 \bmod 3 \]
  and the stronger result $\hat{i}-\hat{j}=0$ follows from the fact
  that $|\hat{i}-\hat{j}|$ is always at most $2$. When $\wh{i-j}\neq0$
  this reasoning breaks down and indeed we have \eg
  $\hat{2}-\hat{1}\neq\wh{2-1}$.}:
\beq\label{add1} \wh{i+j}=\wh{\hat{i}+\hat{j}} \hsc \wh{-i}=-\hat{i} \eeq
\beq\label{add2}  \wh{i-j}=0 \Longleftrightarrow \hat{i}=\hat{j} \eeq
(but note that $\hat{1}+\hat{1}\neq\wh{1+1}$). These are the properties
that will be used in the following. Particularly useful will be the
property
\beq \hat{i}=\hat{j} \Longleftrightarrow \wh{i+k}=\wh{j+k}, \eeq
which follows from (\ref{add2}).

The bracket is manifestly anti-symmetric so to obtain a Lie algebra, there
remains to verify the Jacobi identity. Since
\beq
 [[T_i,T_j],T_k]=\hat{c}_{ijk}T_{i+j+k} \hsc
 c_{ijk}\equiv(i-j)(i+j-k),
\eeq
the Jacobi identity takes the form
\beq\label{jac} \hat{c}_{ijk}+\hat{c}_{jki}+\hat{c}_{kij}=0. \eeq
This identity holds without the `hats', therefore, by (\ref{add1}),
\beq\label{vir}  \hat{c}_{ijk}+\hat{c}_{jki}+\hat{c}_{kij}=0 \bmod 3, \eeq
so (\ref{jac}) can be false only when
\beq  \hat{c}_{ijk}=\hat{c}_{jki}=\hat{c}_{kij}=\pm1. \eeq
$\hat{c}_{ijk}=1$ is equivalent to $\wh{i-j}=\wh{i+j-k}=\pm1$ and,
therefore, also to
\[ i=j\pm1 \bmod 3 \hsc k=-j \bmod 3 \]
and this cannot hold simultaneously for all the tree cyclic permutations
of $\{ijk\}$. Replacing $i\leftrightarrow j$ one obtains the same
result for $\hat{c}_{ijk}=-1$. Therefore, the Jacobi identity holds and
the above algebra is indeed a Lie algebra (over the integers)%
\footnote{In Appendix A.1 we comment about possible generalizations of
  this algebra.}

Let us consider the subalgebra
\[ \Ac_{\infty}\equiv \sp\{T_i\}_{i\geq0}. \]
Dividing by the ideal $\sp\{T_i\}_{i>n}$ (for some positive integer
$n$), one obtains the  finite dimensional Lie algebra
\[ \Ac_n\equiv \sp\{T_i\}_{0\leq i\leq n} \]
with the Lie bracket
\beq
  [T_i,T_j]=\left\{\begin{array}{cc}
    \wh{i-j}T_{i+j}	&	i+j\leq n \\
    0			&	\mbox{otherwise}.
  \end{array}\right.
\eeq
{}From now on we restrict our attention to such an algebra. It is a
solvable%
\footnote{Solvability of a Lie algebra $\Ac$ is defined as follows:
  One defines recursively $\Ac^{k+1}=[\Ac^k,\Ac^k]$; $\Ac$ is
  {\em solvable} iff for some $k$, $\Ac^k=0$.}
algebra, $T_0$ being the only non-nilpotent generator  and it
possesses a $\ZZ$-grading: deg$(T_i)=i$ (inherited from the original
infinite-dimensional algebra.)

We would like to find an invariant metric $(\cdot,\cdot)$ on $\Ac_n$.
Using (\ref{lie}), the invariance condition
\[ ([T_k,T_i] , T_j) + (T_i , [T_k,T_j]) = 0 \]
takes the form
\beq\label{inv} \wh{k-i}(T_{k+i},T_j)+\wh{k-j}(T_{k+j},T_i)=0 \eeq
(here $T_i\equiv0$ for $i>0$) and, in particular, for $k=0$:
\beq \wh{\wh{-i}+\wh{-j}}(T_i,T_j)=0, \eeq
which, by eqs. (\ref{add1},\ref{add2}), is equivalent to
\beq\label{diag} \wh{i+j}(T_i,T_j)=0. \eeq
This means that two out of each three ``reversed''
(right-up-to-left-down) diagonals vanish. Let us look for a metric
with only one non-vanishing diagonal.
To obtain a non-degenerate form, this must be the central diagonal and
according to (\ref{diag}), this is possible only for $\hat{n}=0$. We,
therefore, concentrate on this case and consider a metric of the form
\beq (T_i,T_j)=\om_j\dl_{i+j,n}\hsc \om_{n-j}=\om_j\neq0. \eeq
For such a metric the invariance condition (\ref{inv}) takes the form
\beq \wh{k-i}\om_j+\wh{k-j}\om_i=0, \hs \forall i+j+k=n \eeq
and using $\hat{n}=0$, one obtains
\beq\label{om1} \wh{2i+j}\om_j+\wh{2j+i}\om_i=0. \eeq
First we take $\hat{j}=0$ which gives
\beq \hat{i}(\om_i+\hat{2}\om_j)=0. \eeq
and this implies (since $\hat{2}\neq0$)
\beq \label{om2}
  \om_i=\left\{\begin{array}{cc}
    \om_i=-\hat{2}\om_0	&      \hat{i}\neq0 \\
    \om_i=\om_0        	&      \hat{i}=0.
  \end{array}\right.
\eeq
Using this result we take $\hat{i},\hat{j}\neq0$ in (\ref{om1})
and obtain
\beq \hat{2}\cdot\hat{3}\wh{i+j}\om_0=0, \eeq
which is satisfied, since%
\footnote{This is where the derivation stops being valid for the
  Virasoro algebra (mentioned in Appendix A.1, where possible
  generalizations of the algebra (\ref{Bracket-An}) are discussed).}
$\hat{3}=0$. $-\hat{2}=1$, therefore, we have $\om_i=\om_0,\;\forall i$.
To summarize, we proved:

\noindent{\bf Lemma:}
\begin{quote}\em
  A (non-degenerate) invariant metric on $\Ac_n$ with only one
  (reversed) diagonal exists iff $\hat{n}=0$ and it is proportional to
  \beq (T_i,T_j)=\dl_{i+j-n}. \eeq
\end{quote}
Note that one can add to the metric a multiple of the Killing form,
obtaining
\beq (T_i,T_j)=\dl_{i+j-n}+b\dl_i\dl_j \eeq
(with $b$ arbitrary). The appearance of the second term can also be
seen as a result of the (automorphic) change of basis
\[ T_0\goto T_0+\half bT_n. \]

\subsection{The Ideals in $\Ac_n$}

In this subsection we continue to analyze the algebra $\Ac_n$,
looking for all its ideals and concluding that the
only ideals are of the form
\[ \Ac_{m,n}\equiv \sp\{T_i\}_{i=m}^n. \]
This will be important in the next subsection, where we will check if
these algebras are double extensions of Abelian algebras.
The grading on $\Ac_n$ (deg$(T_i)=i$) will play a central role in the
following and will be called ``charge''. The adjoint action of $T_i$
increases the charge by $i$. Note that there are only positive charges,
so that the adjoint action cannot decrease the charge. This
proves that $\Ac_{m,n}$ (for any $m$) is indeed an ideal.

Let $\Jc$ be an ideal in $\Ac_n$. We choose a basis for $\Jc$ such that
each element has a {\em different} minimal charge (this can be easily
accomplished) and, therefore, can be labeled by it. We, therefore, have
(after an appropriate normalization):
\beq
  \Jc=\sp\{S_{\al}\}, \hs S_{\al}-T_{\al}\in\Ac_{\al+1,n}.
\eeq
Isolating in $\Jc$ the {\em maximal} ideal of the form $\Ac_{m,n}$, we
obtain:
\beq
  \Jc=\sp\{S_{\al}\}_{\al\in\Ab}\bigoplus\Ac_{m,n} \hsc
  m-1\not\in\Ab.
\eeq
Observe that this implies that for any element in $\Jc$ that is not in
$\Ac_{m,n}$, its minimal charge is in $\Ab$.

The choice $\Ab=\emptyset$ (the empty set) corresponds to the ``trivial''
solution $\Jc=\Ac_{m,n}$. In the following we look for other solutions,
\ie with $\Ab\neq\emptyset$. This also implies max$(\Ab)<m-1$.
We are going to explore the restrictions on the $S_{\al}$'s implied by
the claim that $\Jc$ is an ideal in $\Ac_n$. Since $\Ac_{m,n}$ is an ideal
by itself, the only restrictions come from
\beq [T_i,S_{\al}]\in\Jc \hs \forall\al\in\Ab\hsc i=0,\ldots,n. \eeq
$\Jc$ contains all terms with charge at least $m$, therefore, restrictions
will arise only in terms in the commutator with smaller charge. For
$i\geq m-\al$ there are no such terms. As the charge $i$ decreases,
there will be more non-trivial terms, therefore, we will start from the
higher charges.

For $i=m-\al-1$ we have (in the following, ``$\simeq$'' means ``equality
up to an element of $\Ac_{m,n}$''):
\beq
  [T_{m-\al-1},S_{\al}]\simeq
  [T_{m-\al-1},T_{\al}]=\wh{m-2\al-1}T_{m-1}
\eeq
(here and in other similar cases the hat should be applied to the
whole expression between parenthesis).
$T_{m-1}\not\in\Jc$ (otherwise $\Ac_{m-1,n}\subset\Jc$), therefore,
\beq \wh{m-2\al-1}=0. \eeq
Using eqs. (\ref{add1},\ref{add2}), this is equivalent to
\beq\label{halbt} \hat{\al}=-\wh{2\al}=-\wh{m-1} \eeq
and since this is true for all $\al\in\Ab$, we also have
\beq\label{halal}
  \hat{\al}_1=\hat{\al}_2 \hs \forall\al_1,\al_2\in\Ab.
\eeq

Next, for $i=m-\al-2$ we have (using eqs. (\ref{halbt}) and
(\ref{add1}))
\beq\label{second}
  [T_{m-\al-2},S_{\al}]\simeq
  [T_{m-\al-2},T_{\al}+s_{\al}^{\al+1}T_{\al+1}]=
  -T_{m-2}+s_{\al}^{\al+1}T_{m-1}.
\eeq
This implies that $m-2$ is a minimal charge of an element of $\Jc$,
therefore, $m-2\in\Ab$. Substituting $\al=m-2$ in (\ref{second}) we
obtain
\beq
  [T_0,S_{m-2}]\simeq
  -T_{m-2}+s_{m-2}^{m-1}T_{m-1}\simeq
  -S_{m-2}+2s_{m-2}^{m-1}T_{m-1}
\eeq
and this implies $s_{m-2}^{m-1}=0$, so with no loss of generality, we
can choose
\beq S_{m-2}=T_{m-2}. \eeq

Finally, for $i=m-\al-3$ and $m-2>\al\in\Ab$ we have
\beq
  [T_{m-\al-3},S_{\al}]\simeq
  [T_{m-\al-3},T_{\al}+s_{\al}^{\al+1}T_{\al+1}+s_{\al}^{\al+2}T_{\al+2}]=
  T_{m-3}+s_{\al}^{\al+2}T_{m-1}
\eeq
and as before this should imply that $m-3\in\Ab$ (being the minimal
charge of an element of $\Jc$). However, according to eq. (\ref{halal}),
this is impossible since $m-2\in\Ab$. Therefore, $\Ab$ contains no
elements other then $m-2$ and $\Jc$ is of the form
\beq \Jc=\sp\{T_{m-2}\}\oplus\Ac_{m,n}. \eeq
A straightforward check (or use of eq.\ (\ref{halbt})) shows that this
is indeed an ideal iff $\hat{m}=0$. Is this ideal really non-trivial?
It turns out that it is not! To see this, consider the (non-singular)
linear map defined by $T_i\mapsto T'_i\equiv-T_{i+\hat{i}}$. Since
$\hat{m}=0$, this map transforms $\Jc$
to $\Ac_{m-1,n}$.
\beq
  [T'_i,T'_j]=-\wh{i-j}T_{i+j+(\hat{i}+\hat{j})}
             =-\wh{i-j}T_{i+j+\wh{i+j}}=\wh{i-j}T'_{i+j}
\eeq
(the second equality follows from the fact that for $\wh{i-j}\neq0$,
$\wh{i+j}=\hat{i}+\hat{j}$), therefore, this map is an automorphism of
Lie algebras, which means that
$\Jc=\sp\{T_{m-2}\}\oplus\Ac_{m,n}$ is automorphic to $\Ac_{m-1,n}$.

\subsection{$\Ac_n$ as a Deeper algebra}

Now we are ready to check how the self-dual algebras found above fit
into the general picture described in the beginning of this section.
We consider here the case $\hat{n}=0$.
The list of the ideals found in the previous subsection implies that
none of these algebras is decomposable (\ie expressible as an orthogonal
direct sum)%
\footnote{This means that they should be expressible as double
extensions by the one dimensional algebra, and this structure can be
indeed easily identified:
\beq \Bc=\sp\{T_0\} \hsc \Bc^*=\sp\{T_n\} \hsc \Ac=\Ac_{1,n}/\Bc^*. \eeq}.
Among the indecomposable self-dual algebras, we have the following
inclusion relations:
\begin{center}
  $\{$ Indecomposable, Self-Dual Algebras $\}$ \\
  $\cup$ \\
  $\{$ (Single) Double-Extensions of Abelian Algebras $\}$ \\
  $\cup$ \\
  $\{$ Algebras obtainable by a Wigner contraction $\}$
\end{center}
We will show that these are {\em strict} inclusions, \ie, all the three
sets are distinct. Explicitly we will show here that among the algebras
$\Ac_n$, $\Ac_3$ can be
obtained by a Wigner contraction, $\Ac_6$ is a double extension of an
Abelian algebra but cannot be obtained by a Wigner contraction, and the
rest are deeper algebras \ie they are {\em not} double extensions of
Abelian algebras and, therefore, in particular, they {\em cannot} be
obtained by a Wigner contraction.

We start by trying to identify in $\Ac_n$ the structure of a double
extension of an Abelian algebra. The Lie product in an algebra $\Dc$,
with such a structure (table (\ref{LieD})) is of the following form
\beq \begin{array}{c|ccc}
  [\cdot,\cdot] &  \Bc   &  \Ac   &  \Bc^* \\ \hline
  \Bc           &  \Bc   &  \Ac   &  \Bc^* \\
  \Ac           &  \Ac   &  \Bc^* &  0  \\
  \Bc^*         &  \Bc^* &  0     &  0
\end{array}.\eeq
where $\Ac=\sp\{x_i\}$, $\Bc=\sp\{y_a\}$ and $\Bc^*=\sp\{z_a\}$. In
this table we recognize two properties of $\Dc$
\begin{enumerate}
  \item $\Dc$ is a semi direct sum of $\Bc$ and the ideal
    $\Jc=\Ac+\Bc^*$: $\Dc=\Bc\sds\Jc$;
  \item $[\Jc,\Jc]\subset\Bc^*$, therefore,
    $\dim[\Jc,\Jc]\leq\dim\Bc^*=\dim\Bc$.
\end{enumerate}

Consider the first property. The candidates for the ideal
$\Jc$ were found in the previous subsection. It was shown that
$\Jc=\Ac_{m,n}$ (possibly after an automorphic change of basis
$\{T_i\}$). Following the same approach, we choose a basis
$\{R_i\}_{i=0}^{m-1}$ for $\Bc$ such that $i$ is the minimal charge
of $R_i$. $[T_{m-1},T_{m-2}]=T_{2m-3}$ and $2m-3<n$ (since
dim$\Ac_n\geq2$dim$\Bc$), therefore, $[R_{m-1},R_{m-2}]\neq0$ and its
minimal charge is $2m-3$. $\Bc$ is
closed under the Lie bracket and $\Bc\cap\Jc=\{0\}$, therefore,
$[R_{m-1},R_{m-2}]\not\in\Jc$, which implies that $2m-3<m$.
This leaves us with\footnote%
 {The value $m=0$ is also a possibility but it is not interesting.
  It corresponds to $\dim\Bc=0$.
  As a double extension it means not to do anything --
  remaining with the (Abelian) algebra $\Ac$ one started with. As a
  Wigner contraction it means that, starting with some self-dual Lie
  algebra $\Sc_0$, all we did is to set its Lie bracket to $0$, so
  that we end up with the Abelian Lie algebra of the same dimension,
  which is trivially self dual. In the present context, this
  corresponds to the one dimensional algebra:$\Dc=\Ac_0$}
$m=1$ or $2$.

As for the second property, we have
\beq\label{JJm} dim[\Jc,\Jc]\leq\dim\Bc=m. \eeq
One can easily verify that
\beq [\Ac_{m,n},\Ac_{m,n}]=\Ac_{2m+1,n}, \eeq
therefore, eq. (\ref{JJm}) implies $n\leq3m$. On the other hand
$n+1\geq2m$ (since $\dim\Ac_n\geq2\dim\Bc$). Recalling that
$\hat{n}=0$, We obtain three possibilities:
\beq\label{pos} (m,n)=(1,3),(2,3),(2,6) \eeq
and a direct check confirms that each of them indeed corresponds to a
double extension of an Abelian algebra (in the second possibility
this is the zero-dimensional algebra). Observe that there are more
than one way to represent an algebra as a double extension. Moreover,
$\Ac_6$ can be obtained both by extending an Abelian algebra (with
$m=2$) and by extending a non-Abelian algebra (with $m=1$), so the
number of double extensions leading to a given Lie algebra is not
unique%
\footnote{The notion of ``depth'' of a self-dual Lie algebra, suggested
  in \cite{FOF-Stan}, is still well defined, if one allows only extensions
  by either a simple or a one-dimensional algebra. Alternatively, the depth
  can be defined as the {\em minimal} number of double extensions.}.

Turning to the search of the structure of a Wigner contraction, the only
candidates are those enumerated in (\ref{pos}). $\Ac_3$ is the Heisenberg
algebra, and it is indeed a Wigner contraction of $so(2,1)\oplus so(2)$
(which leads to the first possibility in (\ref{pos})). The other candidate
is $\Ac_6$, which corresponds to the last possibility in (\ref{pos}).
To examine this case, we use the further requirement that in a Wigner
contraction, $\Bc$ must be self dual%
\footnote{Actually, the metric is not involved at all in the
  construction of an algebra by a Wigner contraction (unlike double
  extension), and all that is needed is a Lie bracket of the form
  (\ref{LieSa}--\ref{LieSb}).
  However, we are interested in an algebra {\em} with an invariant metric
  and if we want that this procedure will provide us also with the
  metric (through (\ref{metricORS})), $\Bc$ must be self-dual.}.
For $m=2$, $\Bc$ is the two-dimensional, non-Abelian Lie algebra
\[ [R_0,R_1]=R_1. \]
This algebra is not self-dual, therefore, even if $\Ac_6$ can be obtained
by a Wigner contraction, this procedure will not lead to an invariant
metric on $\Ac_6$.

\newsection{Some General Issues Arrising in Gauged WZNW Models Based on
            Non-Semi-Simple Groups}

Having a family of self-dual algebras, the natural thing to do is to
construct the WZNW models based on them. This will be done (for $\Ac_3$ and
$\Ac_6$) in the next section. However, as in any non-compact Lie-algebra,
the invariant metric is not positive definite. In fact, for
all the algebras described in section 2, the metric has more than one
negative eigenvalue, therefore the $\sg$-model obtained from a WZNW model
based on them has an unphysical metric -- more than one time-like direction --
and to correct this we have to gauge out the extra time-like directions.
In the proccess of exploring the various possibilities of
gauging, we encountered some phenomena that are very common when the
algebras involved are not semi-simple. Therefore, before we turn to the
consideration of specific models, we describe in this section some
of these phenomena and analyse them in a general setting. We start, is
subsection 3.1, with a review of
the construction of WZNW and gauged WZNW models. In subsection 3.2 we
consider situations in which the integration of the gauge fields leads to
constraints on the coordinates parametrizing the group manifold. In
subsection 3.3 we discuss ``singular'' gauging, where the restriction of
the metric to the gauged subgroup is degenerate. Finaly, in subsection 3.4
we analyze the gauging of a central subgroup.

\subsection{The General Setup}\label{WZNWgeneral}

To define a WZNW model \cite{WZNW}, one needs a Lie group $G$ and an
invariant metric $(\cdot,\cdot)$ on its Lie algebra $dG$. The
action of the model is%
\footnote{The coupling $k$ is contained here in the metric.}
\footnote{The {\em invariance} of the metric is needed to obtain a
  representation of the affine Lie algebra. It is less apparent
  why the metric should be {\em invertible}. When the group is simple this
  question does not arise, since an invariant (non-zero) metric on a
  simple algebra is always invertible, but this is not true in general.
  In Appendix A.2 we show that by relaxing the condition of invertibility
  one does not obtain any new models, therefore, with no loss of generality,
  we consider only non-degenerate forms.}
\beq\label{SWZNW}
  S[g]=\frac{\hbar}{8\pi}\left[
  \int_\Sg d^2\sg\sqrt{|h|}h^{\al\bt}(J_\al^L,J_\bt^L)
  -\frac{1}{3}\int_Bd^3\sg\ep^{\al\bt\gm}(J_\al^L,[J_\bt^L,J_\gm^L])
  \right],
\eeq
where the field $g$ is a map from a two dimensional manifold $\Sg$ to
$G$, $h_{\al\bt}$ is a metric on $\Sg$, and $J^L=g\inv dg$ is the left
invariant form on $G$ taking values in $dG$. In the second term, $B$ is
an arbitrary three-dimensional manifold such that $\Sg$ is its boundary
and $g$ is extended arbitrarily from $\Sg$ to $B$. Choosing a
parametrization $x^\mu$ for $g$ and substituting it in (\ref{SWZNW})
one obtains (at least locally) a $\sg$-model action (with vanishing dilaton)%
\footnote{The conventional coupling constant $\al'$ is contained here in
  the background fields $G$ and $B$.}
\begin{eqnarray}\label{Ssig}
  S[x] & = & \frac{\hbar}{8\pi}\int_\Sg d^2\sg
             (\sqrt{|h|}h^{\al\bt}G_{\mu\nu}(x)+\ep^{\al\bt}B_{\mu\nu}(x))
             \pt_\al x^\mu\pt_\bt x^\nu \\ \nonumber
       & = & \frac{\hbar}{8\pi}\int_\Sg d^2\sg
             (\sqrt{|h|}h^{\al\bt}+\ep^{\al\bt})E_{\mu\nu}(x)
             \pt_\al x^\mu\pt_\bt x^\nu \hsc
             E_{\mu\nu}\equiv G_{\mu\nu}+B_{\mu\nu}
\end{eqnarray}

The WZNW action (\ref{SWZNW}) is invariant
under the group $G_L\otimes G_R$, acting in $G$ by
\beq\label{hgh} g\goto h_L g h_R\inv \hsc h_L,h_R\in G. \eeq
Given a subgroup $H$ of $G_L\otimes G_R$, one might attempt to {\em gauge}
it, \ie to introduce a $dH$-valued gauge field $A$ and to construct an
extension $\hat{S}$ of (\ref{SWZNW}) that will be invariant under
{\em local} $H$ transformations
\beq\label{ghgh}
  g(\sg)\goto h_L(\sg)g(\sg)h_R(\sg)\inv \hsc
  (h_L(\sg),h_R(\sg))\in H.
\eeq
Such an extension exists 
{\em iff} $H$ is {\em anomaly free}. This criterion can be stated as
follows. Let $H_{L,R}$ be the images of $H$ under the natural
homomorphisms $G_L\otimes G_R\goto G_{L,R}$. These homomorphisms define
corresponding homomorphisms on the algebras:
\beq\label{ALR} dH\goto dH_{L,R} \hsc A\mapsto A^{L,R}. \eeq
The criterion for gauge invariance is that for any $A_1,A_2\in dH$
\beq\label{anomaly} (A^L_1,A^L_2)=(A^R_1,A^R_2). \eeq
Equivalently one can say that the two metrics induced on $dH$ by (the
pullbacks of) (\ref{ALR}) are the same.
Assuming $H$ is indeed anomaly free, the gauge invariant action is
\cite{GWZNW}
\beq\label{SGWZNW}
  \hat{S}[g,A]=S[g]+\frac{\hbar}{4\pi}\int_\Sg d^2\sg
  (\sqrt{|h|}h^{\al\bt}+\ep^{\al\bt})[(A^L_\al,J^R_\bt)
  -(J^L_\al,A^R_\bt)+(A_\al,A_\bt)-(A^L_\al,gA^R_\bt g\inv)]
\eeq
where $J^R=dgg\inv$ is the right invariant form on $G$ and $(A_\al,A_\bt)$
should be understood as $(A^L_\al,A^L_\bt)=(A^R_\al,A^R_\bt)$.
If $h_{\al\bt}$ is conformally flat, we can
choose light-cone coordinates $\sg^\pm$, for which the line element on
$\Sg$ is
\beq ds^2=2e^{2\vph(\sg)}d\sg^+d\sg^- \eeq
which means that
\beq
  \sqrt{|h|}h^{\al\bt}=
  \left(\begin{array}{cc}0 & 1\\ 1 & 0 \end{array}\right).
\eeq
In such coordinates (and with $\ep^{+-}=1$) the action (\ref{SGWZNW}) takes
the simple form
\beq\label{SGWZNWlc}
  \hat{S}[g,A]=S[g]+\frac{\hbar}{2\pi}\int_\Sg d^2\sg
  [(A^L_+,J^R_-)-(J^L_+,A^R_-)+(A_+,A_-)-(A^L_+,gA^R_-g\inv)]
\eeq

To obtain a $\sg$-model description \cite{GWZNW-NLSM}, one integrates
out the gauge fields and fixes the gauge. The action is at most quadratic
in the gauge fields, therefore, the integration can be performed explicitly.
The resulting effective action for $g$ is
\beq\label{SGeff}
  \hat{S}_{\rm eff}[g]=
  \hat{S}[g,A]|_{A=A_{\rm cl}}
  +(\mbox{dilaton term}),
\eeq
where $A_{\rm cl}$ is the solution of the classical
equations for $A$: $\dl\hat{S}/\dl A=0$, and the dilaton term originates
from the functional determinant which arises in the process of integration
\cite{Dilaton}. Another possible contribution to the effective action is
the trace anomaly \cite{Mult-Anomaly}. When $dH$ is self dual
(\eg Abelian), the adjoint representation of $dH$ is traceless,
and there is no trace anomaly. However, when the adjoint representation of
$dH$ is not traceless, such a contribution exists and usually leads to a
non-local action. When the effective action {\em is} local (see also below),
it is a $\sg$-model action
\begin{eqnarray}\label{SGsig}
  \hat{S}_{\rm eff}[x] & = & \frac{\hbar}{8\pi}\int_\Sg d^2\sg
    [(\sqrt{|h|}h^{\al\bt} G_{\mu\nu}(x)
    +\ep^{\al\bt} B_{\mu\nu}(x))\pt_\al x^\mu\pt_\bt x^\nu \\
  \nonumber & & \hspace{30mm}
    +\sqrt{|h|}R^{(2)}\Phi(x)]
\end{eqnarray}
(where $x^\mu$ is an appropriate parametrization for $g$) and in light-cone
coordinates (where $\sqrt{|h|}R^{(2)}=-8\pt_+\pt_-\vph$), it simplifies to
\beq
  \hat{S}[x]=\frac{\hbar}{4\pi}\int_\Sg d^2\sg
  [ E_{\mu\nu}(x)\pt_+x^\mu\pt_-x^\nu
  -4\Phi(x)\pt_+\pt_-\vph].
\eeq

The models presented above are expected to be conformally invariant.
When the action is of the $\sg$-model type, we can verify this to one
loop order, by the vanishing of the beta function equations
\cite{beta}. All the models derived in section 4 passed this check
successfully.

\subsection{The Appearance of Constraints}
\label{constraints}

To obtain a more explicit expression for
$\hat{S}_{\rm eff}$ in (\ref{SGeff}), let us express $A_\pm$
as linear combinations of {\em two distinct} bases $\{T^+_a\}$ and
$\{T^-_a\}$ for $dH$: $A_\pm=A_\pm^aT^\pm_a$. Substituting this into
(\ref{SGWZNWlc}), we obtain
\beq\label{SGWZNcoo}
  \hat{S}[g,A]=S[g]+\frac{\hbar}{2\pi}\int_\Sg d^2\sg
  [A_+^aJ^-_a-J^+_bA_-^b+A_+^aM_{ab}A_-^b]
\eeq
where
\beq J^-_a=(T^{L+}_a,J^R_-) \hsc J^+_b=(J^L_+,T^{R-}_b), \eeq
\beq\label{Mab}
  M_{ab}=(T^{L+}_a,T^{L-}_b-gT^{R-}_bg\inv)
        =(T^{R+}_a-g\inv T^{L+}_ag,T^{R-}_b)
\eeq
and $\{T^{L\pm}_a\}$ and $\{T^{R\pm}_a\}$ are the images of
$\{T^\pm_a\}$ in $dH_L$ and $dH_R$ respectively\footnote%
 {Note that although these sets span $dH_{L,R}$, they are linearly
  independent only if the homomorphisms $dH\goto dH_{L,R}$ are
  injective and we do not assume that it is necessarily so.}.
The classical equations for $A$ are
\beq\label{eqA}
  M_{ab}A_-^b+J^-_a=0 \hsc A_+^aM_{ab}-J^+_b=0.
\eeq
The matrix $M$ is not invertible in general (more about this later).
However, it defines a
bilinear form on $dH$ (a rank-2 covariant tensor under a change of
basis in $dH$) and, therefore, there exists a pair of bases, for which
it is diagonal, and in particular assumes the general form%
\footnote
 {$M$ is not symmetric in general, therefore, the diagonalization
  cannot always be performed with a {\em single} basis. However, in
  some important cases it can. There always exists a (single) basis
  for which $M$ takes the form
  \[ M_{ab}=\left(\begin{array}{cc}
     \bup{M}_{\bup{a}\bup{b}}&0\\ \bdn{M}_{\bdn{a}\bup{b}}&0
     \end{array}\right) \]
  with $\bup{M}_{\bup{a}\bup{b}}$ as above. Therefore, in situations
  where $M_{ab}=0\;\forall g$ implies $M_{ba}=0\;\forall g$, we will
  obtain the form (\ref{Mdiag}). This is what happens in ``vector''
  and ``axial'' gauging: according to (\ref{Mab}) we have (for any gauge)
  \[ M_{ab}=(T^R_b,T^R_a-g\inv T^L_ag) \]
  and for $T^L_a=\pm T^R_a$ this implies
  \[ M_{ab}(g)=M_{ba}(g\inv). \]}
\beq\label{Mdiag}
  M_{ab}=\left(\begin{array}{cc}\bup{M}_{\bup{a}\bup{b}}&0\\0&0
               \end{array}\right),
\eeq
where $\bup{M}_{\bup{a}\bup{b}}$ is a square and invertible matrix
(for a generic choice of $g$) and we divided the set $\{a\}$ of
indices into two sets $\{\bup{a}\}$ and $\{\bdn{a}\}$. In these bases
the equations (\ref{eqA}) are equivalent to
\beq\label{Aeq}
  A_-^\bup{b}=-\bup{N}^{\bup{b}\bup{a}}J^-_\bup{a} \hsc
  A_+^\bup{a}=J^+_\bup{b}\bup{N}^{\bup{b}\bup{a}}.
\eeq
(where $\bup{N}=\bup{M}\inv$:
$\bup{M}_{\bup{a}\bup{b}}\bup{N}^{\bup{b}\bup{c}}=\dl_\bup{a}^\bup{c}$)
and
\beq\label{const} J^-_\bdn{a}=0 \hsc J^+_\bdn{b}=0. \eeq
Substituting this into (\ref{SGWZNcoo}), we obtain
\beq
  \hat{S}[g,A]|_{A=A_{\rm cl}}=
    \{S[g]+\frac{\hbar}{2\pi}\int_\Sg d^2\sg
    [J^+_\bup{b}\bup{N}^{\bup{b}\bup{a}}J^-_\bup{a}]
  \}_{J^-_\bdn{a}=J^+_\bdn{b}=0}.
\eeq
Note that $A_+^\bdn{a}$ and $A_-^\bdn{b}$ remain undetermined
but disappear from the action.

Using an appropriate parametrization $x^\mu$ for $g$, one obtains an
action of the form
\beq
  \hat{S}_{\rm eff}[x]=\frac{\hbar}{4\pi}\int_\Sg d^2\sg
  [ E_{\mu\nu}(x)\pt_+x^\mu\pt_-x^\nu
  -4\Phi(x)\pt_+\pt_-\vph]_{J^-_\bdn{a}=J^+_\bdn{b}=0}.
\eeq
where the dilaton background field is
\begin{eqnarray}
  \Phi(x) & = & -\half\log|\det\bup{M}|+\mbox{const.} \\
          & & +(\mbox{contributions from the delta functions }
                \dl(J^-_{\bdn{a}}),\dl(J^+_{\bdn{b}})).
\end{eqnarray}

This looks almost as a $\sg$-model action, but one must still fix the
gauge and implement the constraints. When there are no constraints,
this indeed leads to a $\sg$-model action, but, since the constraints
are not algebraic, their implementation may lead to a more complicated
(\eg non local) action%
\footnote{In this context it is interesting to know when the constraints
  appear. In Appendix A.3 we show that when an appropriate {\em single}
  ``diagonalizing'' basis $T^+_a=T^-_a$ for $dH$ exists (and in
  particular for axial and vector gauging), constraints can
  appear only if $G$ is not semi-simple or the gauging is singular (\ie
  the metric $(\cdot,\cdot)_{dH}$ induced on $dH$ is degenerate).}.

\subsection{Singular Gauging and Extended Gauge Invariance}
\label{singular}

By ``singular gauging'' we mean gauging a subgroup $H$ such that the
metric induced on its algebra $dH$ is degenerate. This includes the
extreme case of ``null gauging'', for which the metric vanishes
completely. Consider first this last case. The term $(A_+,A_-)$ vanishes
identically, therefore, $A^L_-$ and $A^R_+$ do not appear in the action
(\ref{SGWZNWlc}). This means that the action is determined only by $H_L$
and $H_R$ and there is no trace of the particular choice%
\footnote{The gauge field has the following general form
  \[ A^{L,R}_\pm=A^a_\pm T^{L,R}_a \]
  and $H$ is determined by the relation between $T^L_a$ and $T^R_a$,
  which manifest itself by the dependence between $A^L_+$ and
  $A^R_+$ and similarly between $A^L_-$ and $A^R_-$.}
of a subgroup $H$ of $H_L\otimes H_R$. In particular, the action coincides
with the action obtained by gauging the whole $H_L\otimes H_R$ group.
This has the following important implications:
\begin{enumerate}
  \item The gauge symmetry group of the resulting model is $H_L\otimes H_R$
    and is typically larger than the group $H$ we intended to gauge (the
    group for which we introduced gauge fields). To obtain a $\sg$-model,
    one has to fix this whole extended gauge freedom. This explains why null
    gauging usually reduces the dimension of the $\sg$-model target manifold
    more then by dim$H$ (for $H_L$ and $H_R$ isomorphic to $H$ it
    reduces by 2dim$H$ -- this was observed already in \cite{Null} in a
    specific example).
  \item The vanishing of the metric on $dH$, guaranties that the
    anomaly condition (\ref{anomaly}) is satisfied for each choice of
    a subgroup $H$ of $\hat{H}=H_L\otimes H_R$ and, therefore, one might
    expect that in this case the variety of possible models is considerably
    larger. Contrary to this expectation, we found that all these
    potential models coincide%
\footnote{For $H_L=H_R$ Abelian, this is a trivial realization of
      axial-vector duality \cite{Duality}.}
\footnote{Note that this result holds also when $H$ is central (\ie
      commutes with all elements of $G$), in spite of the fact that
      the diagonal (``vector'') subgroup of $C_L\otimes C_R$, for $C$
      central, acts trivially in $G$ and, therefore, cannot be gauged
      in the usual sense. This point is explained in Appendix A.4.}.
\end{enumerate}

Returning to the general case, we define
\beq \Jc\equiv\{R\in dH|(R,T)=0,\;\forall T\in dH\}. \eeq
The invariance of the metric implies that $\Jc$ is an ideal in $dH$,
therefore, it corresponds to a normal subgroup $N$ of $H$: $\Jc=dN$.
Taking for an element of $H$ a parametrization of the form
$h(x,y)=n(x)k(y)$, where $n\in N$ and $k$ parametrizes elements of
the quotient group, $K\equiv H/N$, the action of $H$ in $G$ is
\beq g\goto[n_L(x)k_L(y)]g[k_R\inv(y)n_R\inv(x)], \eeq
however, similar arguments to those presented above imply that the
action is actually invariant also under local $N_L\otimes N_R$,
\ie $N$ acts independently from the left and from the right,
so the action of the full gauge group is:
\beq g\goto[n_L(x_L)k_L(y)]g[k_R\inv(y)n_R\inv(x_R)] \eeq
(observe that this is indeed a group, \ie closed under composition,
because $N$ is a {\em normal} subgroup of $H$).

\subsection{Gauging a Central Subgroup}
\label{central}

Gauging a central subgroup (\ie taking $H_L$ and $H_R$ that commute with
all elements of $G$) is expected to be a relatively simple choice of
gauging.  However, in many cases such a choice does not lead
to a new $\sg$-model backgrounds. In particular, we show here that when
$H_L=H_R$, the resulting model is an (ungauged) WZNW model%
\footnote{In Appendix A.5 we analyze the $H_L\neq H_R$ case for one
  dimensional $H$. We also comment on the $H_L\neq H_R$ case of null $H$
  in a footnote after eq. (\ref{Snull}). These cases also do not lead to
  new $\sg$-model backgrounds.}.
Denoting%
\footnote{We note the distinction between $\bar{H}\subset G$ and
  $H\subset H_L\otimes H_R\subset G_L\otimes G_R$.}
$\bar{H}=H_L=H_R$, the model is based on the group $K/\bar{H}_0$, where $K$
is the subgroup of $G$ generated by all the generators of $G$ that are
orthogonal to $d\bar{H}$, and $\bar{H}_0$ is the subgroup of $\bar{H}$
generated by all the ``null''  generators: $d\bar{H}_0=d\bar{H}\cap dK$.

First we show that those generators of $G$ that are in $d\bar{H}$ but not
in $d\bar{H}_0$ do not contribute to the final gauged action.
The restriction of the metric on $dG$ to
$d\bar{H}$, being a symmetric bilinear form on $d\bar{H}$, can be
diagonalized, therefore, if it does not vanish, we have a
decomposition $d\bar{H}=d\bar{H}_0\oplus d\bar{H}_1$, such that
$(d\bar{H},d\bar{H}_0)=0$ and the restriction of the metric to
$d\bar{H}_1$ is non-degenerate. Using the invariance of the metric
on $dG$, we have
\beq\label{ggz} ([dG,dG],d\bar{H}_1)=(dG,[dG,d\bar{H}_1])=0 \eeq
(since $\bar{H}$ is central), which implies that $[dG,dG]$ is contained
in $d\bar{H}_1^\perp$ and, in particular, that $d\bar{H}_1^\perp$ is a
subalgebra of $dG$ (in fact it is an ideal, since it commutes with
$d\bar{H}_1$). Let $G_0$ be the corresponding subgroup of $G$:
$(d\bar{H}_1)^\perp=dG_0$. The metric on $d\bar{H}_1$ is non-degenerate,
therefore,
\beq dG_0\cap d\bar{H}_1=\{0\}, \eeq
so $dG$ is a direct sum of orthogonal ideals
\beq\label{dGds} dG=dG_0\oplus d\bar{H}_1 \hsc (dG_0,d\bar{H}_1)=0. \eeq
The action of a WZNW model based on a direct product of groups
$G=G_0\otimes G_1$, where the algebras of the groups are orthogonal to
each other, decomposes to a sum of independent terms, a term for each
factor. When the gauged group is also a direct product of the form
\beq\label{Hdp}
  H=H_0\otimes H_1 \hsc H_i\subset (G_i)_L\otimes (G_i)_R \hsc
  (i=0,1)
\eeq
the decomposition of the action holds also in the gauged model.
However, since $\bar{H}_0$ is null, (\ref{Hdp}) always holds.
To show this, let us construct an appropriate basis for $dH$. Since
$dH\subset dH_L\oplus dH_R$, a basis element is represented by a pair
$((T^L_i)_0+(T^L_i)_1,(T^R_i)_0+(T^R_i)_1)$. According to the general
discussion of null gauging, $dH_0=d(H_0)_L\oplus d(H_0)_R$, so we can
choose a basis (for $dH$) that includes a basis for $d(H_0)_L$
(\ie $(T^L_i)_1=(T^R_i)_0=(T^R_i)_1)=0$) and a basis for $d(H_0)_R$.
With such a basis we can set $(T^L_i)_0=(T^R_i)_0=0$ whenever
{\em either} $(T^L_i)_1\neq0$ {\em or} $(T^R_i)_1\neq0$ and the
resulting set will remain a basis. This means that we indeed have the
direct product structure (\ref{Hdp})%
\footnote{This result depends on the assumption $H_L=H_R$. See
  Appendix A.5 for a counter-example in the general case.},
with $G_1=\bar{H}_1$ and we can analyze each factor separately.
The $G_1$ part corresponds to $G$ Abelian, $H_L=H_R=G$, and its
contribution to the action vanishes (more about this in Appendix A.4).
This enables us (up to topological issues) to restrict our attension
to the $G_0$ part of $G$ and the $H_0$ part of $H$, \ie, to consider
null $H$.

According to (\ref{ggz}),
\beq\label{GGinH} [dG,dG]\subset d\bar{H}^\perp, \eeq
where $d\bar{H}^\perp$ is the orthogonal complement of $d\bar{H}$. This
implies that $d\bar{H}^\perp$ is a subalgebra. We will denote by $K$
the corresponding subgroup of $G$. Note that $\bar{H}\subset K$, since
$d\bar{H}$ is null. To construct a parametrization of $G$ we chose a
subspace $\Bc$ of $dG$ such that
$dG=\Bc\oplus dK$ (direct sum of vector spaces) and a basis
$\{S_a\}_{a=1}^m$ for $\Bc$ ($m=\dim\Bc=dim\bar{H}$). We use
(\ref{GGinH}) again, to deduce that $dG$ has the following structure%
\footnote{\label{sdsp} Recall that we denote by $\Bc\sds\Jc$ a
  semi-direct sum of algebras $\Bc$ and $\Jc$, in which $\Jc$ is an
  ideal. Similarly, $H\sdp N$ denotes a semi-direct product of
  groups $H$ and $N$, in which $N$ is a normal subgroup.}
\beq\label{RRK}
  dG=\RR S_1\sds(\RR S_2\sds(\ldots\sds(\RR S_m\sds dK)\ldots))
\eeq
(where $\RR S_a\equiv\sp_{\RR}\{S_a\}$) and, therefore%
\footnote{For some Abelian subalgebra of $dG$ contained in $\Bc$,
  the corresponding subgroup of $G$ may be compact. In this case the
  lefthand side of (\ref{GRG})  should be divided by a discrete group.
  This is irrelevant to the subsequent discussion and, therefore, will
  be ignored.},
\cite{Hau-Sch}
\beq\label{GRG}
  G=e^{\RR S_1}\sdp(e^{\RR S_2}\sdp
  (\ldots\sdp(e^{\RR S_m}\sdp K)\ldots))
\eeq
(where $\exp(\RR S_a)$ denotes the one-parameter-group generated by
$S_a$). This suggest a parametrization for $g\in G$ of the form
\beq g(x,y,z)=e^{y^mS_m}\ldots e^{y^1S_1}k(x)e^{z^aT_a}, \eeq
where $\{T_a\}_{a=1}^m$ is a basis for $\bar{H}$ and $k(x)$ is some
parametrization of the quotient group $K/\bar{H}$.
In these coordinates, the invariant form $J^L$ takes the form (using
(\ref{sts}))
\begin{eqnarray}
  J^L\equiv g\inv dg & = &
      T_adz^a+k\inv dk+k\inv[S_1dy^1+e^{-y^1\ad(S_1)}(S_2dy^2+ \\
  \nonumber & & +e^{-y^2\ad(S_2)}(\ldots(S_{m-1}dy^{m-1}
      +e^{-y^{m-1}\ad(S_{m-1})}S_mdy^m)\ldots))]k \\
  \nonumber & = & T_adz^a+S_ady^a+J^L_\perp(x,y) \hsc J^L_\perp\in dK
\end{eqnarray}
(the last equality follows from $e^STe^{-S}-T\in[dG,dG]\subset dK$),
so the gauged WZNW action is of the form
\beq\label{Snull}
  \hat{S}[g,A]=S[g]+\frac{\hbar}{2\pi}\int_\Sg d^2\sg
  [(A^L_+,S_a)\pt_-y^a-(A^R_-,S_a)\pt_+y^a].
\eeq
The integration over $A$ yields the constraints
\begin{description}
  \item $\pt_-y^a=0$ \hs for components coupled to $dH_L$;
  \item $\pt_+y^a=0$ \hs for components coupled to $dH_R$.
\end{description}
which means, since we assume%
\footnote{For $dH_L\neq dH_R$ we may define
  \beq\label{Hbar}
    \bar{H}=H_LH_R\equiv\{h_Lh_R|h_{L,R}\in H_{L,R}\}
  \eeq
  which is also a central subgroup. If $\bar{H}$ is null, the whole
  derivation up to this point is valid. $A^L_+$ and $A^R_-$ in eq.
  (\ref{Snull}) are components of {\em two different} gauge fields,
  corresponding to left and right translations, respectively.
  Since $dH_L\neq dH_R$, some components of $y$ will be constrained
  to depend on one of the light-cone coordinates and it is not clear
  if the resulting effective action can be brought to the form of a
  $\sg$-model.}
$H_L=H_R$, that $dy^a=0\;\forall a$. $J^L$ simplifies to
\beq
  J^L=T_adz^a+g_0\inv dg_0\in dG_0,
\eeq
the $z$ dependence disappear from the action (because $T_a$ is central
and orthogonal to $dG_0$) and the resulting effective action coincides
with the WZNW action for the group $K/\bar{H}$ (with the target space
variables $x^i$)%
\footnote{Note that all the non-invariant fields disappear from the
  action so no gauge fixing is needed.}.

To summarize, when $H_L=H_R$ is central,
there is always an (orthogonal) direct product decomposition
\beq G=G_0\otimes G_1 \hsc dG_0\perp dG_1, \eeq
\beq
  H=H_0\otimes H_1 \hsc
  (H_0)_L=(H_0)_R\subset G_0 \hsc (H_1)_L=(H_1)_R=G_1
\eeq
such that $G_1$ is central and $H_0$ is null.  Denoting
$\bar{H}_0\equiv (H_0)_L=(H_0)_R$ and $dK=(d\bar{H}_0)^\perp$ we have
\beq G_0\supset K\supset \bar{H}_0 \eeq
and the resulting action is the action of a WZNW model for the group
$K/\bar{H}_0$.

\newsection{WZNW and gauged WZNW models based on $\Ac_n$}

In this section we will present some (gauged and ungauged) WZNW models
based on the algebras $\Ac_3$ and $\Ac_6$. The corresponding $\sg$-model
background fields, and some related tensors, are listed in Appendix B.
For all the $\sg$-models obtained, the one-loop beta functions \cite{beta}
vanish and the central charge is equal to the dimension of the $\sg$-model
target manifold.

For the construction of a WZNW model, one needs a
convenient parametrization of the corresponding group. For a compact
group $G$ one often chooses the parametrization $g=\exp(x^aT_a)$, where
$\{T_a\}$ is some basis for $dG$. However, in general the exponential map
exp: $dG\goto G$ is not onto, even if $G$ is connected%
\footnote{For example, it can be shown that an $SL(2,\RR)$ matrix of the
  form
  \[ \left(\begin{array}{cc} -1 & \bt \\ 0 & -1 \end{array}\right)
     \hsc \bt\neq0
  \]
  is not in the range of the exponential map.}
and when it is not, a different parametrization is needed. For solvable
Lie algebras (as the algebras $\Ac_n$), one can exploit the fact
\cite{Hau-Sch} that any such algebra is a repeated semi-direct
sum of one-dimensional Lie algebras and consequently
\cite{Hau-Sch} the corresponding (connected and simply connected
covering) group is a semi-direct product of one-dimensional groups.
This is the approach we use. The grading on $\Ac_n$ implies that
\beq
  \Ac_n=\RR T_0\sds(\RR T_1\sds(\ldots\sds(\RR T_{n-1}\sds\RR T_n)\ldots))
\eeq
so the simply connected covering group of $\Ac_n$ is
\beq\label{Gn}
  G_n=e^{\RR T_0}\sdp(e^{\RR T_1}\sdp
  (\ldots\sdp(e^{\RR T_{n-1}}\sdp e^{\RR T_n})\ldots)).
\eeq
This means that the map
\beq\label{gpar}
  (x_0,\ldots,x_n)\goto g=e^{x_n T_n}\ldots e^{x_0 T_0}
\eeq
is a homeomorphism from $\RR^{n+1}$ onto $G_n$. This is, therefore, a
suitable parametrization.

Next we derive expressions for the invariant forms on $G_n$. Denoting by
$\ad(S)$ the adjoint action of $S\in dG$:
\beq \ad(S):T\goto[S,T] \hsc S,T\in dG \eeq
and
\beq\label{expt}
  e^{x\ad(S)}\equiv\sum_{k=0}^\infty \frac{x^k}{k!}\ad(S)^k,
\eeq
we have
\beq\label{sts} e^S T e^{-S}=e^{\ad(S)}(T). \eeq
Using this formula we obtain:
\begin{eqnarray}\label{JL}
  J^L & \equiv & g\inv dg \\ \nonumber
      & = & T_0dx_0+e^{-x_0\ad(T_0)}(T_1dx_1+e^{-x_1\ad(T_1)}(\ldots
            (T_{n-1}dx_{n-1}+e^{-x_{n-1}\ad(T_{n-1})}T_ndx_n)\ldots)),
\end{eqnarray}
\begin{eqnarray}\label{JR}
  J^R & \equiv & dgg\inv \\ \nonumber
      & = & T_ndx_n+e^{x_n\ad(T_n)}
            (T_{n-1}dx_{n-1}+e^{x_{n-1}\ad(T_{n-1})}(\ldots
            (T_1dx_1+e^{x_1\ad(T_1)}T_0dx_0)\ldots)).
\end{eqnarray}
$T_0$ acts by multiplication:
\beq e^{x_0\ad(T_0)}:T_i\goto e^{-\hat{i}x_0}T_i \eeq
and all other generators are nilpotent, which means that the sum in
(\ref{expt}) is finite, therefore, formulas (\ref{JL}) and (\ref{JR})
provide a well defined algorithm for  the computation of the invariant
forms.

\subsection{Models based on $\Ac_3$}

\subsubsection{The ungauged model}

Using the parametrization (\ref{gpar}) and the formulas
(\ref{JL}) and (\ref{JR}) with $n=3$ we obtain
\begin{eqnarray}
  J^L & = & T_0dx_0+e^{x_0}T_1dx_1+e^{-x_0}T_2dx_2+T_3(x_1dx_2+dx_3) \\
  J^R & = & T_0dx_0+T_1(dx_1+x_1dx_0)+T_2(dx_2-x_2dx_0) \\
  \nonumber & & +T_3(dx_3+x_2dx_1+x_2x_1dx_0)
\end{eqnarray}
Substituting these expressions in the WZNW action (\ref{SWZNW}),
together with the Lie bracket (\ref{lie}) and the invariant
metric%
\footnote{This metric is obtained from the diagonal one (with $b=0$)
  by the (automorphic) change of basis $T_0\goto T_0+\half bT_3$,
  therefore, keeping $b$ arbitrary will provide us with a convenient
  way of performing a family of gaugings.}
\footnote{We could take a constant multiple of (\ref{metric-three}).
  This is equivalent to  changing the $\sg$-model coupling constant
  $\al'$, \ie rescaling the background fields ($G_{\mu\nu}$ and
  $B_{\mu\nu}$ but not $\Phi$).}
\beq\label{metric-three} (T_i,T_j)=\dl_{i+j-3}+b\dl_i\dl_j, \eeq
one obtains the following $\sg$-model action:
\beq\label{Sa3}
  S=\frac{\hbar}{4\pi}\int_\Sg d^2\sg
    [\pt_+x_0(2\pt_-x_3+b\pt_-x_0)+2\pt_+x_2(\pt_-x_1+x_1\pt_-x_0)].
\eeq
This is an analytic continuation of the model in \cite{Nappi-Witten}.

\subsubsection{Gauging $T_0$}

Next we gauge the symmetry
\beq g\goto h_Lgh_R\inv, \hs h_{L,R}=e^{\th_{L,R}T_0} \eeq
which, in the coordinates (\ref{gpar}) takes the form
\begin{eqnarray}
  x_0 & \goto & x_0+\th_L-\th_R \\ \nonumber
  x_1 & \goto & e^{-\th_L}x_1 \\ \nonumber
  x_2 & \goto & e^{\th_L}x_2 \\ \nonumber
  x_3 & \goto & x_3.
\end{eqnarray}
Since $(T_0,T_0)=b$, for $b\neq0$ the anomaly condition is
$\th_L=\pm\th_R$ (``vector/axial'' gauging), while for $b=0$  the
gauging is null, therefore, anomaly-free and, by the general discussion
in section \ref{singular}, independent of the relation between $\th_L$
and $\th_R$. Hence we can restrict attention to the
cases $\th_L=\pm\th_R$. Using the general notation introduced
in section \ref{constraints}, we have
\beq M\equiv(T_0, T_0\mp gT_0g\inv)=b(1\mp1)\mp x_1x_2 \eeq
(since $gT_0g\inv=T_0+x_1T_1-x_2T_2+x_1x_2T_3$),
\beq J^+=(J^L_+,\pm T_0)=\pm(b\pt_+x_0+x_1\pt_+x_2+\pt_+x_3) \eeq
and
\beq J^-=(T_0,J^R_-)=(b+x_1x_2)\pt_-x_0+x_2\pt_-x_1+\pt_-x_3. \eeq
$M$ does not vanish identically (not even for $b=0$), so the effect of
the gauging is to add to the Lagrangian the contribution
\[  \frac{\hbar}{2\pi}\left[\frac{1}{M}J^+J^-+
    (\log M)\pt_+\pt_-\vph\right].
\]

To proceed, we need a gauge choice.
In the case of axial gauging $\th_L=-\th_R$, we can fix $x_0=0$ and
obtain
\begin{eqnarray}\nonumber
  \hat{S}_{\rm eff}(x_1,x_2,x_3) & = &
    \frac{\hbar}{4\pi}\int_\Sg d^2\sg[2\pt_+x_1\pt_-x_2-
    2p(x_1\pt_+x_2+\pt_+x_3)(x_2\pt_-x_1+\pt_-x_3) \\
  \label{Sa3g0a} & & \hspace{25mm} +2\log|x_1x_2+2b|\pt_+\pt_-\vph]
\end{eqnarray}
with
\beq p=\frac{1}{x_1x_2+2b}. \eeq
For vector gauging $\th_L=\th_R$, $x_0$ is invariant and the symmetry
acts only on $x_1$ and $x_2$. This action does not alter the sign so
strictly speaking one cannot fix a coordinate to a constant. However,
at $x_1x_2=0$ there is a singularity, so  non-singular field
configurations will have coordinates with a homogeneous sign and the
field configuration space is divided to sectors. Moreover, the action
is invariant%
\footnote{This invariance originates from the fact that the map
  $T_i\goto-T_i$, $i=1,2$, is an isometric automorphism in $\Ac_3$,
  \ie it conserves the metric and the Lie bracket.}
under $x_i\goto-x_i$, $i=1,2$ so we can restrict ourselves to the
$x_1>0$ sector. Therefore, we {\em can} choose the gauge $x_1=1$ and
the result is
\begin{eqnarray}\label{Sa3g0v}
  \hat{S}_{\rm eff}(x_0,x_2,x_3) & = & \frac{\hbar}{4\pi}
    \int_\Sg d^2\sg[\pt_+x_0\pt_-(bx_0+2x_2+2x_3)- \\
  \nonumber & & -\frac{2}{x_2}\pt_+(bx_0+x_2+x_3)
    ((b+x_2)\pt_-x_0+\pt_-x_3)+2\log|x_2|\pt_+\pt_-\vph].
\end{eqnarray}

For $b=0$ the two models in eqs. (\ref{Sa3g0a}) and (\ref{Sa3g0v})
are apparently different, although they should be the same according to
the general discussion in section \ref{singular}. But this is the result
of a different gauge choice. In fact the gauge choice $x_1=1$ is equally
valid for the axial gauging when $b=0$ and it leads to identical models.
The metric in eqs. (\ref{Sa3g0a}) and (\ref{Sa3g0v}) with $b=0$
is degenerate (see Appendix B). This is expected, since $b=0$
corresponds to a null gauging, and the degeneracy is the result of
the extended gauge symmetry, as discussed in section \ref{singular}.
Indeed, the independence of $\th_L$ and
$\th_R$ allows the fixing of both $x_0$ and $x_1$. The action obtained is
\beq\label{Sa3g0n}
  \hat{S}_{\rm eff}(x_2,x_3)=\frac{\hbar}{4\pi}\int_\Sg d^2\sg
  [-\frac{2}{x_2}\pt_+(x_2+x_3)\pt_-x_3
  +2\log|x_2|\pt_+\pt_-\vph],
\eeq
and the corresponding metric is non-degenerate.

\subsection{Models based on $\Ac_6$}

\subsubsection{The ungauged model}

Using the parametrization (\ref{gpar}) and the formulas (\ref{JL}) and
(\ref{JR}) with $n=6$ we obtain
\begin{eqnarray}
  J^L & = & T_0dx_0+T_1e^{x_0}dx_1
            +T_2e^{-x_0}dx_2+T_3(dx_3+x_1dx_2) \\ \nonumber
      &   & +T_4e^{x_0}(dx_4-x_1dx_3-\half x_1^2dx_2)
            +T_5e^{-x_0}(dx_5+x_2dx_3)               \\  \nonumber
      &   & +T_6(dx_6+x_1dx_5-x_2dx_4+x_1x_2dx_3)
\end{eqnarray}
and
\begin{eqnarray}
  J^R & = & T_0dx_0+T_1(dx_1+x_1dx_0)+T_2(dx_2-x_2dx_0) \\ \nonumber
      &   & +T_3(dx_3+x_2(dx_1+x_1dx_0))
            +T_4(dx_4-x_3dx_1+(x_4-x_3x_1)dx_0) \\ \nonumber
      &   & +T_5(dx_5+x_3dx_2-\half x_2^2dx_1
                 -(x_5+x_3x_2+\half x_2^2x_1)dx_0) \\ \nonumber
      &   & +T_6(dx_6-x_4dx_2+x_5dx_1+(x_5x_1+x_4x_2)dx_0).
\end{eqnarray}
Substituting these expressions in the WZNW action (\ref{SWZNW}),
together with the Lie bracket (\ref{lie}) and the metric
\beq\label{metric-six} (T_i,T_j)=\dl_{i+j-6}+b\dl_i\dl_j, \eeq
one obtains the following $\sg$-model action:
\begin{eqnarray}\label{Sa6}
  S & = & \frac{\hbar}{4\pi}\int_\Sg d^2\sg
          [\sum_{i=0}^6\pt_+x_i\pt_-x_{6-i}+b\pt_+x_0\pt_-x_0+ \\
  \nonumber & & \hspace{20mm}
           +x_1(\pt_+x_2\pt_-x_3-\pt_+x_3\pt_-x_2+2\pt_+x_5\pt_-x_0) \\
  \nonumber & & \hspace{20mm}
           +x_2(\pt_+x_1\pt_-x_3+\pt_+x_3\pt_-x_1-2\pt_+x_4\pt_-x_0)
           +2x_1x_2\pt_+x_3\pt_-x_0]
\end{eqnarray}

\subsubsection{Gauging the Center}

The simplest way to obtain a model with a reduced dimension is to gauge
the center of%
\footnote{Recall that $G_6$ is the simply connected covering group of
  $\Ac_6$, as defined in eq. (\ref{Gn}).}
$G_6$, which is the one-parameter group generated by
$T_6$. This will also serve as an illustration of the general discussion
in section \ref{central}. More precisely, we gauge the two-dimensional
subgroup $H$ of $G_L\otimes G_R$ whose action in $G$ is%
\footnote
 {As explained in the general discussion, this is equivalent to
  vector/axial gauging of a one dimensional subgroup, which in our
  notation corresponds to $\th_L=\pm\th_R$.}
\beq g\goto h_Lgh_R\inv, \hs h_{L,R}=e^{\th_{L,R}T_6} \eeq
(in the parametrization (\ref{gpar}), this is
$x_6\goto x_6+\th_L-\th_R$). This gauging is anomaly-free
for independent $\th_L$ and $\th_R$ because $T_6$ is null in the
metric (\ref{metric-six}). The resulting action is
\beq
  \hat{S}[g,A]=S[g]+\frac{\hbar}{2\pi}\int_\Sg d^2\sg
  (A_+\pt_--A_-\pt_+)x_0.
\eeq
Integrating out the gauge fields results with the constraint
$x_0=$const.\ . Imposing the constraint in the action $\hat{S}$
eliminates the $x_0$ and $x_6$ dependence and we obtain a 5-dimensional
(gauge-invariant) $\sg$-model action:
\begin{eqnarray}\label{Sa6g6}
  \lefteqn{\hat{S}_{\rm eff}(x_1,x_2,x_3,x_4,x_5)=S|_{dx_0=0}} \\
    \nonumber & & =\frac{\hbar}{4\pi}\int_\Sg d^2\sg
                [\sum_{i=1}^5\pt_+x_i\pt_-x_{6-i}
                +x_1(\pt_+x_2\pt_-x_3-\pt_+x_3\pt_-x_2) \\
    \nonumber & & \hspace{25mm}
                +x_2(\pt_+x_1\pt_-x_3+\pt_+x_3\pt_-x_1)].
\end{eqnarray}
which is the WZNW action for the group whose algebra is%
\footnote{This is the unique 5-dimensional self-dual Lie algebra,
  appearing in the list of \cite{Kehagias}.}
\beq\label{Aa16g6} \Ac=\sp\{T_1,\ldots,T_6\}/\sp\{T_6\}. \eeq

\subsubsection{Towards a Four Dimensional Model}

To get down to 4 dimensions, we must add another generator to $H$.
We want to explore as many options as possible, so we take the
action of $H$ in $G$ to be
\beq\label{httf}
  g\goto h_Lgh_R\inv \hsc h_{L,R}=e^{\th_{L,R}T_6+\vph_{L,R}T_m}
\eeq
with $T_m=a_2 T_2+a_4 T_4$, where $a_{2,4}$ are parameters which
determine the choice of the additional generator. Since
\beq (T_m,T_6)=0 \hsc (T_m,T_m)=2a_2a_4, \eeq
for $a_2a_4\neq0$ the anomaly condition is $\vph_L=\pm\vph_R$
(``vector/axial'' gauging), while for $a_2a_4=0$ ($T_m$ proportional to
$T_2$ or $T_4$) the gauging is null and, therefore, anomaly-free for any
$\vph_{L,R}$. To fit to the general notation introduced in section
(\ref{WZNWgeneral}), we choose a (single) basis for $dH$ with%
\footnote{More precisely, the basis that corresponds to the action
  (\ref{httf}) is
  \[ \{T^L_a\}=\{T_m,T_6,0,0\} \hsc  \{T^R_a\}=\{0,0,T_m,T_6\}, \]
  where the first two generators generate the left action and the other
  two generate the right action.
  The choice (\ref{tms}) corresponds to vectorial gauging of the central
  element: $\th_L=\th_R$ and to some left-right correlated gauging of
  $T_m$: $\vph_{L,R}=\al_{L,R}\vph$. The last restriction is actually
  necessary when $T_m$ is not null, while for null generators ($T_0$
  and possibly also $T_m$) we have seen that the resulting action is
  unaffected by these restrictions. Therefore, there is no loss of
  generality in the choice (\ref{tms}).}
\beq\label{tms}
  \{T^L_a\}=\{\al_LT_m,T_6\} \hsc  \{T^R_a\}=\{\al_RT_m,T_6\}.
\eeq
\begin{eqnarray}
  gT_mg\inv & = & a_2e^{x_0}(T_2-x_1T_3+(x_1x_2+x_3)T_5) \\ \nonumber
  & & +(a_4e^{-x_0}-\half a_2e^{x_0}x_1^2)T_4
      +[a_4x_2e^{-x_0}-a_2e^{x_0}(x_4+\half x_1^2x_2)]T_6,
\end{eqnarray}
therefore,
\beq
  M_{ab}\equiv(T^L_a,T^L_b-gT^R_bg\inv)=
  \left(\begin{array}{cc}M&0\\0&0\end{array}\right)
\eeq
with
\beq
  M=\half a_2^2\al_L\al_Re^{x_0}x_1^2
  +2a_2a_4(\al^2-\al_L\al_R\cosh x_0)
\eeq
(recall that when $a_2a_4\neq0$, $\al_L^2=\al_R^2\equiv\al^2$).
{}From the $T_6$ gauging we obtain, as before, the constraints $dx_0=0$.
As to the other generator, the corresponding current components are
(for $dx_0=0$)
\begin{eqnarray}
  J^+_m = (J^L_+,\al_RT_m) & = &
    \al_R[(a_4e^{-x_0}-\half a_2e^{x_0}x_1^2)\pt_+x_2 \\
  \nonumber & & +a_2e^{x_0}(-x_1\pt_+x_3+\pt_+x_4)] \\
  J^-_m = (\al_LT_m,J^R_-) & = &
    \al_L[-a_2x_3\pt_-x_1+a_4\pt_-x_2+a_2\pt_-x_4]
\end{eqnarray}
and we have several different situations:
\begin{enumerate}
  \item $a_2=0$:

    This implies $M=0$, which leads to the constraints $J^+_m=J^-_m=0$.
    For $\al_L\al_R\neq0$ this is equivalent to $dx_2=0$. Imposing these
    constraints in the action $\hat{S}$ eliminates the $x_4$ dependence
    and we obtain a 3-dimensional $\sg$-model action:
    \begin{eqnarray}\label{Stfour}
      \hat{S}_{\rm eff}(x_1,x_3,x_5) & = & S|_{dx_0=dx_2=0} = \\
      \nonumber & = & \frac{\hbar}{4\pi}\int_\Sg d^2\sg
        [2\pt_+x_5\pt_-x_1+\pt_+x_3\pt_-x_3+2x_2\pt_+x_3\pt_-x_1]
    \end{eqnarray}
    which corresponds to a constant (flat) background:
    \beq
       E_{\mu\nu}=\left(\begin{array}{ccc}
      0   & x_2 & 1 \\
      x_2 & 1   & 0 \\
      1   & 0   & 0
      \end{array}\right).
    \eeq

  \item $\al_L\al_R=0$:

    This requires $a_2a_4=0$ ($T_m$ null) and, therefore, also in this
    case $M$ vanishes. But in the present case either $T^L_m$ or $T^R_m$
    vanish, therefore, we obtain only {\em one} constraint:
    \begin{eqnarray}
      a_2=0: && \al_L\pt_-x_2=\al_R\pt_+x_2=0 \\
      a_4=0: && \al_L(\pt_-x_4-x_3\pt_-x_1)= \\ \nonumber
             && =\al_R(\pt_+x_4-x_1\pt_+x_3-\half x_1^2\pt_+x_2)=0
    \end{eqnarray}
    which eliminates roughly ``half'' a degree of freedom, therefore, it
    seems that the resulting model is not of a $\sg$-model type.

  \item $a_2,\al_L\al_R\neq0$:

    (note that this includes the null case $a_4=0$)

    In this case $M$ does not vanish identically and the effect of the
    gauging (after integrating out the gauge field) is to add to the
    WZNW Lagrangian the contribution
    \[ \frac{\hbar}{2\pi}\left[\frac{1}{M}J^+_mJ^-_m
       +(\log M)\pt_+\pt_-\vph\right].
    \]
\end{enumerate}

\subsubsection{A Four Dimensional Model}

We continue with the last case. To choose an appropriate gauge fixing
condition, we need the explicit action of $H$ in $G$. Using repeatedly
a generalization of eq. (\ref{sts}):
\beq e^Sf(T)e^{-S}=f\left(e^{\ad(S)}(T)\right) \eeq
(valid for any function $f$ expressible as a
convergent power series) and the Campbell-Baker-Hausdorff formula
\cite{Hau-Sch}, which for $[S,T]$ that commutes with $S$ and $T$ takes
the form
\beq e^Se^T=s^{S+T+\half[S,T]}, \eeq
we obtain
\begin{eqnarray}
  \nonumber x_0 & \goto & x_0 \\
  \nonumber x_1 & \goto & x_1 \\
  \nonumber x_2 & \goto & x_2+a_2(\vph_L-e^{x_0}\vph_R) \\
  \label{coo-trans} x_3 & \goto & x_3+a_2e^{x_0}x_1\vph_R \\
  \nonumber x_4 & \goto & x_4+a_4(\vph_L-e^{-x_0}\vph_R)
                +\half a_2e^{x_0}x_1^2\vph_R \\
  \nonumber x_5 & \goto & x_5-a_2(x_3\vph_L+e^{x_0}x_1x_2\vph_R)
     +a_2^2e^{x_0}x_1(\half e^{x_0}\vph_R-\vph_L)\vph_R \\
  \nonumber x_6 & \goto & x_6+(\th_L-\th_R)
     +a_2x_4\vph_L+(\half a_2e^{x_0}x_1^2-a_4e^{-x_0})x_2\vph_R \\
  \nonumber & & +\half a_2^2e^{x_0}x_1^2(\vph_L
     -\half e^{x_0}\vph_R)\vph_R+a_2a_4(\vph^2-e^{-x_0}\vph_L\vph_R).
\end{eqnarray}
Since we chose $a_2\neq0$, one may fix $x_2$ as a gauge condition,
unless $e^{x_0}=\al_L/\al_R$. At this stage we should observe that
$x_0$ is not a fixed parameter of the theory but rather an integration
variable -- it is the remnant of the $[dx_0]$ functional integration
(in the partition function), which was not fixed by the constraint
$dx_0=0$. A single isolated value of $x_0$ has measure zero and does
not have any influence on the integral. Therefore, we may restrict
ourselves to the generic case
\beq e^{x_0}\neq\al_L/\al_R. \eeq
We choose the gauge $x_2=$const., impose the constraint $x_0=$const.,
and obtain the  $\sg$-model action
\begin{eqnarray}\label{Sa6g642}
  \hat{S}_{\rm eff}(x_1,x_3,x_4,x_5) & = &
    \frac{\hbar}{4\pi}\int_\Sg d^2\sg
    [2\pt_+x_5\pt_-x_1+\pt_+x_3\pt_-x_3+2x_2\pt_+x_3\pt_-x_1 \\
  \nonumber & & +4p(x_1\pt_+x_3-\pt_+x_4)(x_3\pt_-x_1-\pt_-x_4)
    +2\log(x_1^2+a)\pt_+\pt_-\vph]
\end{eqnarray}
with
\beq p=\frac{\al_L\al_R a_2^2}{2M}e^{x_0}=\frac{1}{x_1^2+a} \eeq
\beq
  a=4\frac{a_4}{a_2}e^{-x_0}\left(\frac{\al_L}{\al_R}-\cosh x_0\right)
   =-2\frac{a_4}{a_2}\left(\frac{\al_L}{\al_R}-e^{-x_0}\right)^2
\eeq
(we used the fact that when $a_4\neq0$, the anomaly condition imposes
the constraint $\al_L/\al_R=\pm1$). Note that all the parameters that
determine the $T_m$-gauge are concentrated in one parameter $a$. Its
sign is equal to the sign of $-(T_m,T_m)$ (since the other
factors are strictly positive by assumption). We now show that the
magnitude of $a$ carries no physical information, therefore, only its
sign is important. Indeed, the action (\ref{Sa6g642}) is
invariant under the transformation%
\footnote{This transformation corresponds to a change of basis
  $T_i\goto\lm^{-\hat{i}}T_i$ in $\Ac_n$. For $\lm>0$ this is the
  adjoint action of $\lm^{T_0}\in G_n$ in $\Ac_n$, which is always an (
  inner) isometric automorphism (\ie conserves the metric and the Lie
  bracket). For $\lm=-1$ a simple check shows that it is also an
  isometric automorphism (although outer). This explains the invariance
  of the action (although it can be also easily verified directly).}
\beq
  x_i\goto\lm^{\hat{i}}x_i,\;a\goto\lm^2a  \hs \lm\neq0
\eeq
(where $\hat{i}\equiv i\bmod3\in\{-1,0,1\}$),
therefore, a change in the magnitude of $a$ is equivalent to a
coordinate transformation. Furthermore, for positive $\lm$ the
coordinate transformation (this time keeping $a$ unchanged!) is a
result of the adjoint action of $\lm^{-T_0}$ in $G_n$
\beq g\goto\lm^{T_0}g\lm^{-T_0}. \eeq
The functional measure $[dg]$ is invariant under such a transformation,
therefore, the partition function is independent of the magnitude of $a$.
This has several important implications:
\begin{enumerate}
  \item The action is essentially (as an integrand) independent of the
    value of $x_0$ and the $dx_0$ integration in the partition function
    is trivial%
\footnote{We did not consider the value $e^{x_0}=\al_L/\al_R$ but, as
      explained already, this is a single point in the $dx_0$ integral
      and, therefore, can not have any influence.}.
    Therefore, we {\em can} view $x_0$ as a fixed parameter in
    the action (void of any physical content) and not as an integration
    variable and the resulting effective action is indeed of the
    $\sg$-model type.
  \item The model is independent of the choice of $\al_L$ and $\al_R$
    (as long as they don't vanish). In other words, we again have
    trivial vectorial/axial duality (and its generalization in the null
    case).
  \item The model is almost independent of the choice of $a_2$ and
    $a_4$, which determine the direction of $T_m$ in the ($T_2,T_4$)
    plane. Only the sign of $-a_2a_4$ (which is the signature of $T_m$)
    is significant.
\end{enumerate}
The fact that $x_0$ can be treated as a parameter implies that the model
(\ref{Sa6g642}) coincides with the WZNW model based on the 5-dimensional
algebra (\ref{Aa16g6}) gauged by sp$\{T_m\}$. This model was derived in
\cite{Kehagias}, using a different basis for the algebra and a different
parametrization of the group manifold. The action obtained using those
choices belongs to a family of exactly conformal $\sg$-model actions of
the form
\beq\label{Tseytlin}
  S[u,v,y_i]=\frac{\hbar}{4\pi}\int_\Sg d^2\sg
  \{k[2\pt_+v\pt_-u+U_{ij}(u)\pt_+y_i\pt_-y_j]-4\Phi(u)\pt_+\pt_-\vph\},
\eeq
considered in \cite{Exact-compare}. Moreover, the action corresponding to
the axial gauging was obtained in \cite{Exact-compare} as a special limit
of the $[E_2^c\otimes U(1)]/U(1)$ (vectorially) gauged WZNW model.
It was also shown in \cite{Kehagias} that the actions obtained are related
to 4-dimensional flat actions by duality \cite{Duality}.

Incidentally, observe that in the limit $a_2/a_4\goto0$, $p\goto0$ and
we obtain the action (\ref{Stfour}) of the $a_2=0$ case. In both cases
we have $x_2=$const., however, for $a_2\neq0$ this is a {\em gauge
choice} and the model is explicitly independent of the value chosen for
$x_2$. On the other hand, for $a_2=0$, $x_2$ is {\em constrained} to be
a constant and one still must {\em apriori} integrate over its
value. However, the fact that the $a_2=0$ model can be obtained as a
limit of the $a_2\neq0$ model implies that the $a_2=0$ model is also
independent of $x_2$ and the integration is unnecessary. In the present
model this is trivially seen directly, but in more complicated models,
where the independence may not be obvious, this may be a convenient way
to prove it.

For $a=0$ ($a_4=0$) the metric of the model (\ref{Sa6g642}) is
degenerate, but as in the $\Ac_3$ models, this is because in this case
$T_m$ is null (proportional to $T_2$), and we have an extended gauge
symmetry ($\vph_L,\vph_R$ independent in (\ref{httf})). According to
(\ref{coo-trans}) (with $a_4=0$), for $x_1\neq0$ one can choose the
gauge $x_2=x_3=0$ and the resulting action is
\begin{eqnarray}\label{Sa6g62}
  \hat{S}_{\rm eff}(x_1,x_4,x_5) & = &
    \frac{\hbar}{4\pi}\int_\Sg d^2\sg[2\pt_+x_5\pt_-x_1
    +\frac{4}{x_1^2}\pt_+x_4\pt_-x_4+4\log(x_1)\pt_+\pt_-\vph].
\end{eqnarray}

\subsubsection{Non-Abelian gauging}

In the search for a $\sg$-model with a physical signature, one may consider
also non-Abelian subalgebras. However, since $\Ac_n$ is solvable, its two and
three dimensional subalgebras are not self-dual and, therefore, one expects
a contribution from the trace anomaly \cite{Mult-Anomaly}. We considered
two examples, $dH=\sp\{T_0,T_2\}$ and $dH=\sp\{T_0,T_5\}$. Both of them are
of the type $[T,S]=S$. The trace of (the adjoint representation of) $T$ is
non-trivial, so the trace anomaly contributes and the $\sg$-model
background fields calculated according to the formulas in section 3 are
expected not to satisfy the beta function equations. This is indeed what
happens for the first model ($dH=\sp\{T_0,T_2\}$). The second model leads to
a constant (flat) $E_{\mu\nu}$ and a linear dilaton. The one-loop beta
functions for such a background vanish with a shift in the central charge
(relative to the dimension of the $\sg$-model target manifold). We expect
the trace anomaly contribution to cancel the dilaton, leading to a flat
background.

\newsection{Summary and Remarks}
In this work we investigated WZNW and gauged WZNW models based on
non-reductive algebras. We introduced a family $\{\Ac_{3m}\}$ of
such algebras that are not double extensions of Abelian algebras and,
therefore, cannot be obtained through a Wigner contraction.
This may provide one with a new family of conformal field theories.

We constructed WZNW and gauged WZNW models based on the first two
algebras in this series: $\Ac_3$ and $\Ac_6$. The purpose was to
find models that can serve as string vacua, and also to gain general
knowledge about the use of non-reductive algebras (and the family
$\Ac_n$ in particular) in this context. This indeed provided some
general observations, which lead to the derivation of some general
results concerning singular and central gauging. Here we describe some of
the features and problems of the use of non-reductive algebras for the
construction of WZNW and gauged WZNW models:
\begin{itemize}
 \item
  The $\sg$-model obtained from a WZNW model based on a non-reductive
  algebra
  is never positive definite. Moreover, in the process of constructing
  a non-reductive self-dual algebra, starting from an Abelian one, each
  double extension adds at least one timelike direction. This implies
  that (indecomposable) non-reductive algebras, that are not double
  extensions of Abelian algebras, always lead to an unphysical signature,
  with more than one time-like direction, so to obtain a useful model
  using these algebras, one must gauge out the extra time-like
  directions (see, for example \cite{Sfet-gauge}).
 \item
  Many of the possibilities of gauging a non-reductive
  WZNW model are singular. That singular subalgebras are quite common can be
  seen in the $\Ac_n$ algebras and is also suggested by the structure of a
  double extension. In fact, lower dimensional self dual subalgebras are
  quite rare, as can be seen in the list of \cite{Kehagias}: two dimensional
  non-Abelian sub-algebras are never self-dual and if the (total) algebra is
  solvable (and, therefore, does not contain simple subalgebras), this is true
  also for three dimensional subalgebras.
 \item
  When the gauging is singular the gauge symmetry group of the model is
  typically larger then the subgroup initially chosen to be gauged and as a
  result such a gauging reduces the dimension of the $\sg$-model target
  manifold by more than the dimension of that subgroup. For example, when
  $H_L$ and $H_R$ are one dimensional and null, The dimension is reduced
  by {\em two}.
 \item
  The signature of a $\sg$-model obtained from a gauged WZNW seems to
  be related to the one obtained from the ungauged model in a simple way:
  \begin{itemize}
   \item
    when the gauging is non-singular (\ie, the metric induced on the
    algebra of the gauged subgroup is non-degenerate), the signature of
    the gauged model is obtained from the ungauged one by ``subtracting''
    the signature of the gauged subalgebra;
   \item
    when the gauging is null (and leads to a $\sg$-model) the dimension of
    the target manifold reduces in pairs, consisting of one positive and
    one negative direction.
  \end{itemize}
  Therefore, the choice of a gauged subgroup is restricted to subgroups
  with the signature dictated by the desired final signature.
  In particular, this limits the use of null directions. In the $\Ac_n$
  algebras, for example, the difference between the number of positive
  and negative eigenvalues of the metric is 0 or 1, and to get a
  4-dimensional (or larger) Minkowskian-signature background, one must
  gauge a non-null sub-algebra.
 \item
  Singular gauging leads frequently (although not always) to the
  appearance of constraints which, in many cases, lead to a model
  that is not of a $\sg$-model type. All possible situations were
  demonstrated in the $\Ac_6$ models described in section 4:
  \begin{itemize}
   \item $dH_L=\sp\{T_4\}$, $dH_R=0$ (or vise versa) leads to
    constraints that lead to a non-$\sg$-model system;
   \item $dH_L=dH_R=\sp\{T_4\}$ leads also to constraints but
    these {\em do} lead to a $\sg$-model;
   \item $dH_L=dH_R=\sp\{T_2\}$ does not lead to constraints.
  \end{itemize}
  Another complication is that in singular gaugings, the gauged subgroup
  is not necessarily self-dual and when it is not, there may be a
  non-trivial contribution from the trace anomaly \cite{Mult-Anomaly}
  (as demonstrated at the end of section 4).
  However, there are singularly gauged WZNW models that {\em do} lead to
  a good $\sg$-model, so this possibility does not have to be completely
  avoided.
 \item
  Gauging a central subgroup does not lead to new $\sg$-models, at least for
  $H_L=H_R$. It either leads to an ungauged WZNW model or does not lead
  to a clearly defined field theory at all. For example, gauging the $T_6$
  direction of $\Ac_6$ lead to the WZNW model for
  $\Ac=\sp\{T_1,\ldots,T_6\}/\sp\{T_6\}$.
  We also gauged a subalgebra of $\Ac_6$ that contained the central
  element $T_6$, and the result coincided with a gauged WZNW model based on
  $\Ac$. This seems as a general result. Therefore, to obtain
  genuine new models, it seems that one should avoid subgroups
  containing central elements. If this is true, it limits considerably
  the useful choices of gauging.
\end{itemize}

In spite of the problems and limitations encountered, gauged WZNW models
based on the algebras $\{\Ac_n\}$ and on non-reductive self-dual algebras
in general, may lead to new and interesting string backgrounds (as was already
demonstrated) and therefore deserve further study. In particular, one may
try to derive string backgrounds using the next algebras
of the family $\Ac_n$. These are not double extensions of Abelian
algebras and it would be interesting to see if this property is reflected
in some way in the WZNW models (or in other models based on non-reductive
self-dual algebras \cite{Tseytlin-9505}).
%

Finally, we should remark that an open direction for research is the
construction of the conformal field theory that corresponds to the gauged
WZNW model. For non-singular vectorial gauging this is the coset construction
(for a non-reductive self-dual algebra, this was shown in
\cite{FOF}), but for the other gauging possibilities the corresponding CFT
is not known. In particular it is not clear what is the resulting central
charge. For non-singular vectorial gauging it is equal to the difference
between the central charges of WZNW models based on the group $G$ and on the
subgroup $H$ respectively. When $G$ is solvable, this implies that the
central charge is equal to the dimension of the $\sg$-model target manifold.
All the $\sg$-models derived in section 4 obeyed this rule. It remains to be
seen if this is true in general.

\vspace{1cm}
\noindent{\bf Acknowledgment}

We would like to thank A.A.\ Tseytlin for useful discussions.
The work of AG is supported in part by BSF -- American-Israel Bi-National
Science Foundation, by the BRF -- the Basic Research Foundation and by an
Alon fellowship. The work of ER is supported in part by BSF and by the BRF.

\newpage

\appendix
\renewcommand{\newsection}[1]{
 \vspace{10mm} \pagebreak[3]
 \addtocounter{section}{1}
 \setcounter{equation}{0}
 \setcounter{subsection}{0}
 \setcounter{paragraph}{0}
 \setcounter{equation}{0}
 \setcounter{figure}{0}
 \setcounter{table}{0}
 \addcontentsline{toc}{section}{
  Appendix \protect\numberline{\Alph{section}}{#1}}
 \begin{flushleft}
  {\large\bf Appendix \thesection. \hspace{5mm} #1}
 \end{flushleft}
 \nopagebreak}


\newsection{Comments and Supplements}

\subsection{Generalizations of the algebras $\Ac_n$}

Here we comment about possible generalizations of the algebras defined
in section \ref{An}, obtained by using the defining relations
(\ref{Bracket-An}) with a {\em different} choice of the map
``$\hat{\hspace{3mm}}$''. If one takes ``$\hat{\hspace{3mm}}$'' to
be some homomorphism from $\ZZ$ to some commutative ring $\FF$ with
unity, (\ref{mult}-\ref{add2}) hold, as well as (\ref{jac}) and one
obtains a Lie algebra over $\FF$. For example, one can take $\FF=\ZZ_p$
($p$ a positive integer) with ``$\hat{\hspace{3mm}}$'' being the
natural homomorphism. This example, however, is irrelevant for WZNW
models, since one needs there an algebra over $\RR$ and for this $\FF$
must be some subring of $\RR$. A more relevant example will be obtained
by taking $\FF=\ZZ$ and ``$\hat{\hspace{3mm}}$'' the identity map. The
result is the Virasoro algebra (with zero central charge). Another
natural candidate for ``$\hat{\hspace{3mm}}$'' would be
$\hat{i}=i \bmod p$ ($p$ a positive integer). Taking $p=2$ and
$\hat{\hspace{3mm}}:\ZZ\goto\{0,1\}$, an analysis similar to the $p=3$
case leads to the choice $(i,j,k)=(1,0,0)$, for which the
right-hand-side of (\ref{jac}) does not vanish
($\hat{c}_{ijk}=\hat{c}_{kij}=1$, $\hat{c}_{jki}=0$). There seems to
be no other choice of $p$ and range of the map ``$\hat{\hspace{3mm}}$''
such that the multiplication is preserved. In the main text we
concentrate on the specific choice $\hat{i}=i \bmod 3\in\{-1,0,1\}$,
but we rarely use more then the properties (\ref{mult}-\ref{add2}) and
the Jacobi identity (\ref{jac}), so most of the analysis applies to
possible future generalizations.

\subsection{A WZNW model with a degenerate metric}

Here we analyze the WZNW model (section \ref{WZNWgeneral}) obtained
when one uses a degenerate metric. Let us define
\[ \Jc\equiv\{R\in dG|(R,T)=0,\;\forall T\in dG\}. \]
The invariance of the metric implies that this is an ideal, therefore,
it corresponds to a normal subgroup $N$ of $G$: $\Jc=dN$; and the
metric $(\cdot,\cdot)$ is an invertible metric on the algebra $dG_0$ of
the quotient group $G_0\equiv G/N$. Taking for an element in $G$ a
parametrization of the form $g=ng_0$, where $n\in N$ and $g_0$
parametrizes elements of the quotient group, it is straightforward to
show that the action is independent of $n$, therefore, it actually
corresponds to the group $G_0$ (for which the metric is non-degenerate).
Therefore, relaxing the requirement of non-degeneracy gives us nothing
new.

\subsection{Conditions for the appearance of constraints}

Here we show that for a wide class of gauged WZNW models, constraints
do not appear and, therefore, the resulting model is a non linear
$\sg$-model. We find that when an appropriate {\em single}
``diagonalizing'' basis $T^+_a=T^-_a$ for $dH$ exists (and in particular
for axial and vector gauging), constraints can appear only if $G$
is not semi-simple or the gauging is singular (\ie the metric
$(\cdot,\cdot)_{dH}$ induced on $dH$ is degenerate). As we saw in section
\ref{constraints}, the
appearance of constraints is equivalent to the degeneracy of the matrix
$M_{ab}$, which means that there exist $T^\pm_{\bdn{a}}\in dH$ for which
\beq\label{perp}
  T^{L-}_{\bdn{a}}-gT^{R-}_{\bdn{a}}g\inv\perp dH_L \hsc
  T^{R+}_{\bdn{a}}-g\inv T^{L+}_{\bdn{a}}g\perp dH_R.
\eeq
Assuming $T^+_{\bdn{a}}=T^-_{\bdn{a}}$, this implies
\beq T^L_{\bdn{a}}-T^R_{\bdn{a}}\perp dH_L+dH_R \eeq
and when $(\cdot,\cdot)_{dH}$ is non-degenerate, this implies
$T^L_{\bdn{a}}=T^R_{\bdn{a}}$. Putting this in (\ref{perp}), we obtain
\beq gT_{\bdn{a}}g\inv-T_{\bdn{a}}\perp dH_L+dH_R, \eeq
which means that the space
\beq \Jc\equiv\sp\{gT_{\bdn{a}}g\inv-T_{\bdn{a}}\}_{g\in G} \eeq
is orthogonal to $dH_L+dH_R$. Using relations (\ref{expt}) and
(\ref{sts}), we obtain, for any $S\in dG$
\beq
  [S,gT_{\bdn{a}}g\inv-T_{\bdn{a}}]=\lim_{t\goto0}\frac{1}{t}\left\{
  \left[\left(e^{tS}g\right)T_{\bdn{a}}\left(e^{tS}g\right)\inv
  -T_{\bdn{a}}\right]-\left[e^{tS}T_{\bdn{a}}e^{-tS}-T_{\bdn{a}}\right]
  -\left[gT_{\bdn{a}}g\inv-T_{\bdn{a}}\right]\right\}
\eeq
which means that $\Jc$ is an ideal in $dG$. If $G$ is semi-simple,
$T_{\bdn{a}}$ is not central, therefore, $\Jc$ is not empty (since it
contains $[dG,T_{\bdn{a}}]$), so $\Jc$ is a semi-simple factor:
\beq\label{JJ} dG=\Jc\oplus\Jc' \hsc \Jc\perp\Jc' \eeq
and $dH_L+dH_R$ is contained in the other factor $\Jc'$. This implies
that $T_{\bdn{a}}$ is in $\Jc'$ and, therefore, so is $gT_{\bdn{a}}g\inv$
(since $\Jc$ is an ideal). This would imply that $\Jc$ is contained in
$\Jc'$, in contradiction to (\ref{JJ}) and the non-triviality of $\Jc$.
Thus in this case $T_{\bdn{a}}$ as above cannot exist.

\subsection{Vectorial gauging of a central group}

When the center $Z$ of $G$
\beq Z=\{c\in G|cg=gc\;\forall g\in G\} \eeq
is non-trivial, then $Z$, embedded diagonally in $G_L\otimes G_R$, acts
trivially in $G$:
\beq hgh\inv=g\;\forall h\in Z\;,g\in G \eeq
and the faithfully acting symmetry group of the action (\ref{SWZNW})
is $G_L\otimes G_R/Z$. Therefore, it seems meaningless to ``vectorially
gauge'' the center (or a subgroup of it) since the original action is
already (trivially) gauge invariant. In spite of this, the gauged
action (\ref{SGWZNW}) is different from the ungauged one%
\footnote{The difference should be separately gauge invariant, and
  indeed it is - as can be directly verified.}:
\beq
  \hat{S}[g,A]=S[g]
  +\frac{\hbar}{2\pi}\int_\Sg d^2\sg\ep^{\al\bt}(A_\al,J^L_\bt).
\eeq
Moreover, we showed in section \ref{singular} that when the gauged
group is null, ``vectorial'' gauging is identical to ``axial'' gauging,
where the $g$ and $S$ are not gauge invariant. The question that
arises is, therefore, what does it mean, in this context, to gauge
vectorially a central group.

The situation becomes more clear when we recall that the purpose of
the introduction of the gauge fields in a WZNW model is not to obtain
a gauge invariant theory but rather to obtain a model with a reduced
dimension. The simplest example is that of ``axial'', non-singular
gauging of a central group $C$. We have shown in section \ref{central}
that in such a situation $dG$ is an orthogonal direct sum of ideals
$dG_0\oplus dC$ and, therefore, $G$ has a parametrization
$g=g_0(x)e^{z^iT_i}$ where $g_0(x)$ is some parametrization of $G_0$
and $\{T_i\}$ is a basis for $dC$. The symmetry gauged is
\beq g\goto e^{2\th^iT_i}g(=hhg=hgh) \eeq
which is equivalent to $z^i\goto z^i+2\th^i$. After fixing the gauge
(and integrating out the gauge fields) we are left with a WZNW model
for $G_0$.

In the vectorial case the situation is quite different but
the final result is identical:
keeping in mind the real purpose of the gauging procedure, we add the
additional terms to the (already gauge-invariant) ungauged action. Since
$g$ is gauge-invariant, the dimension can not be reduced by
gauge fixing. Instead, the elimination of physical degrees of freedom
occurs here because of the appearance of constraints, introduced by
the additional terms. Indeed, we have
\beq
  \hat{S}[g,A]=S[g]+\frac{\hbar}{2\pi}\int_\Sg (A,T_i)\wedge dz^i,
\eeq
therefore, the integration of the gauge fields leads to the constraints
$dz^i=0$, which results, again, with a WZNW model for $G_0$.
To summarize, we have shown that ``vectorial gauging'' of a central
group has a well defined meaning in the present context, although the
name ``gauging'' is misleading. We also saw that in all cases of
central gauging (singular or not), vector/axial duality is not only
valid, but also trivial.

\subsection{Gauging a one dimensional central group with $H_L\neq H_R$}
In section \ref{central} we analyzed central gauging when $H_L=H_R$. To
get an idea what new behavior can be expected, when $H_L$ and $H_R$ are
different, we examine here the one dimensional case.

We define (as in (\ref{Hbar})) the group
\beq \bar{H}=H_LH_R\equiv\{h_Lh_R|h_{L,R}\in H_{L,R}\}, \eeq
(which is also a central subgroup of $G$) and analyze different situations
according to the rank of the metric on
\beq d\bar{H}=\sp\{T_L,T_R\}. \eeq
\begin{itemize}
 \item{$\bar{H}$ is null ($(\cdot,\cdot)_{d\bar{H}}$=0)}

   This case was analyzed (for arbitrary dim$\bar{H}$) in a footnote after
   eq. (\ref{Snull}). It contains the case of ``one sided gauging''
   (when one of the groups $H_L$ or $H_R$ vanishes). The effect of the
   gauging is that some of the coordinates will be constrained to depend
   on one of the light-cone coordinates and it is not clear if the
   resulting action can be brought to the form of a $\sg$-model.

 \item{$(\cdot,\cdot)_{d\bar{H}}$ is non-degenerate}

   This implies that
   \beq dG=dG_0\oplus d\bar{H} \hsc dG_0\perp d\bar{H}, \eeq
   which is of the form (\ref{Hdp}) and the result is:
   \begin{itemize}
    \item when $H$ is null: a WZNW model for $G_0$.
    \item when $H$ is not null:

      a WZNW model for $G_0$, tensored with a one dimensional free model.
   \end{itemize}

 \item{rank$(\cdot,\cdot)_{d\bar{H}}$=1}

   This is the only case essentially different from the $H_L=H_R$ ones.
   We choose a diagonalizing basis for $d\bar{H}$
   \beq d\bar{H}=\sp\{T_0,T_1\} \hsc (T_i,T_j)=i\dl_{ij} \eeq
   (note that because of the anomaly condition, $\bar{H}$ is
   necessarily two dimensional in this case) and obtain the structure
   (\ref{dGds}) with $H_i=\sp\{T_i\}$. The anomaly condition implies
   \beq T_L=\al T_1+\bt_L T_0 \hsc T_R=\pm(\al T_1+\bt_R T_0) \eeq
   and $H_L\neq H_R$ implies $\bt_L\neq\bt_R$ and $\al\neq0$.
   We observe that $T_0^\perp$ (the subspace of $dG_0$ orthogonal to
   $T_0$) is a subalgebra containing $T_0$, therefore, corresponds to a
   subgroup of $G_0$, which we denote by $K$.
   Next we choose some element $S$ of $dG_0$ obeying $(S,T_0)=1$ and
   obtain the structure
   \beq dG_0=\sp\{S\}\sds dK \eeq
   which suggests the following parametrization for $g\in G$
   \beq g(x,y,z_0,z_1)=k(x)e^{yS+Z_iT_i}, \eeq
   where k(x) is some parametrization of the quotient group
   $K/\sp\{T_0\}$. In this pa\-ra\-met\-ri\-za\-tion, the invariant form
   $J^L$ is
   \beq
     J^L=T_idz_i+Sdy+J^L_\perp \hsc
     J^L_\perp(x,y)=e^{-yS}k\inv dke^{yS}\in dK
   \eeq
   and this leads, in the notation of section \ref{constraints} to
   \beq
     M=\al^2(1\mp1) \hsc
     J^+=\pm\pt_+(\al z_1+\bt_Ly) \hsc
     J^-=\pt_-(\al z_1+\bt_Ry).
   \eeq
   The coordinate transformation representing the action of $H$
   \beq g\goto e^{\th T_L}ge^{-\th T_R} \eeq
   is
   \beq \dl z_0=\th(\bt_L\mp\bt_R) \hsc \dl z_1=\th\al(1\mp1). \eeq
   For the vector-like gauging (the upper sign), $M=0$ and one obtains
   (gauge-invariant) constraints $J^+=J^-=0$ that seem not to lead to a
   $\sg$-model. For the axial-like gauging (the lower sign), $M\neq0$
   and the effective action is
   \beq
     \hat{S}_{\rm eff}=
     S[g]+\frac{\hbar}{2\pi}\int_\Sg d^2\sg\frac{J^+J^-}{M}.
   \eeq
   With the gauge choice $z_1=0$, $S[g]$ becomes the WZNW action for
   $G_0$ and the effect of the second term is to change the $\sg$-model
   metric
   \beq ds^2\goto ds^2-\frac{\bt_L \bt_R}{\al^2}dy^2, \eeq
   which is equivalent to a change in the value of the norm (S,S). This
   value can be changed by the automorphic redefinition
   $S\goto S+\gm T_0$, therefore,
   the change does not effect the invariance and non-degeneracy of the
   metric. We conclude that the resulting model is a WZNW model for
   $G_0$ with a modified invariant metric.
\end{itemize}
To summarize, we found that in all the cases of one dimensional
central gauging, one either encounters constraints that seem not to lead to
a $\sg$-model, or obtains an ungauged WZNW model.

\newsection{The $\sg$ model backgrounds obtained}

In section 4 we derived several models of the $\sg$-model type:
\begin{eqnarray}
  \nonumber S[x] & = & \frac{\hbar}{8\pi}\int_\Sg d^2\sg
    [(\sqrt{|h|}h^{\al\bt}G_{\mu\nu}+\ep^{\al\bt}B_{\mu\nu}(x))
    \pt_\al x^\mu\pt_\bt x^\nu+\sqrt{|h|}R^{(2)}\Phi(x)] \\
  & = & \frac{\hbar}{4\pi}\int_\Sg d^2\sg
    [E_{\mu\nu}(x)\pt_+x^\mu\pt_-x^\nu-4\Phi(x)\pt_+\pt_-\vph],
\end{eqnarray}
\[ E_{\mu\nu}=G_{\mu\nu}+B_{\mu\nu}. \]
In this Appendix we list, for each model, the corresponding background
fields: the metric $G_{\mu\nu}$, the anti-symmetric tensor $B_{\mu\nu}$
and the dilaton $\Phi$, and some related quantities (indexes are lowered
by $G_{\mu\nu}$ and raised by the inverse metric $G^{\mu\nu}$):
\begin{itemize}
  \item the connection
    \[ {\Gm_{\mu\nu}}^\rho=\half G^{\rho\sg}
      (\pt_\mu G_{\nu\sg}+\pt_\nu G_{\mu\sg}-\pt_\sg G_{\mu\nu});
    \]
  \item the Riemann tensor
    \[ {R_{\mu\nu\rho}}^\sg=2({\Gm_{\rho[\mu}}^\lm{\Gm_{\nu]\lm}}^\sg
      -\pt_{[\mu}{\Gm_{\nu]\rho}}^\sg);
    \]
  \item the Ricci tensor $R_{\mu\nu}={R_{\mu\rho\nu}}^\rho$
    and the Ricci scalar $R=R_{\mu\nu}G^{\mu\nu}$;
  \item the torsion of the anti-symmetric tensor
    \[ H=3dB=3\pt_{[\rho}B_{\mu\nu]}dx^\rho\wedge dx^\mu\wedge dx^\nu;
    \]
  \item contractions of the squared torsion
    \[ H^2_{\mu\nu}=H_{\mu\rho\sg}{H_\nu}^{\rho\sg} \hsc
       H^2=H^2_{\mu\nu}G^{\mu\nu}.
    \]
\end{itemize}
These quantities are needed, among other things, to verify the
one-loop beta function equations \cite{beta}
\begin{eqnarray}\label{bt-Phi}
  0 = \frac{16\pi^2}{\al'}\bt^\Phi & = & \frac{c-d}{3\al'}
    +4(\del\Phi)^2-4\del^2\Phi-R+\frac{1}{12}H^2+\Oc(\al') \\
  0 = \bt^G_{\mu\nu} & = & R_{\mu\nu}-\frac{1}{4}H^2_{\mu\nu}
    +2\del_\mu\del_\nu\Phi+\Oc(\al') \\
  \label{bt-B} 0 = \bt^B_{\mu\nu} & = & \del^\rho H_{\rho\mu\nu}
    -2(\del^\rho\Phi)H_{\rho\mu\nu}+\Oc(\al').
\end{eqnarray}
These equations were indeed found to be satisfied for all the models
presented, with the central charge $c$ equal to the dimension $d$ of
the target manifold.

\subsection{$\Ac_3$ ungauged}
(eq. (\ref{Sa3}))

\noindent The coordinates of the target manifold are
\[ \{x_0,x_1,x_2,x_3\}; \]
the background fields are:
\beq
  G_{\mu\nu}=\left(\begin{array}{cccc}
  b   & 0 & x_1 & 1 \\
  0   & 0 & 1   & 0 \\
  x_1 & 1 & 0   & 0 \\
  1   & 0 & 0   & 0
  \end{array}\right) \hsc
  B_{\mu\nu}=\left(\begin{array}{cccc}
  0   & 0 & -x_1 & 0 \\
  0   & 0 & 0    & 0 \\
  x_1 & 0 & 0    & 0 \\
  0   & 0 & 0    & 0
  \end{array}\right),
\eeq
\[ \Phi=0 \]
($B=2x_1dx_2\wedge dx_0$; the metric with signature $(+,+,-,-)$, as the
metric on $\Ac_3$);

\noindent the inverse metric is
\beq
  G^{\mu\nu}=\left(\begin{array}{cccc}
  0 & 0    & 0 & 1    \\
  0 & 0    & 1 & -x_1 \\
  0 & 1    & 0 & 0    \\
  1 & -x_1 & 0 & -b
  \end{array}\right);
\eeq
the only non-vanishing component of the Riemann tensor is (up to symmetry)
$R_{0120}=\frac{1}{4}$, the only non-vanishing component of the
Ricci tensor is $R_{00}=-\half$ and the Ricci scalar vanishes.

\noindent The torsion of the anti-symmetric tensor is
\beq H=6dx_0\wedge dx_1\wedge dx_2 \eeq
(which means $H_{012}=1$) and it is covariantly constant;

\noindent the only non-vanishing component of $H^2_{\mu\nu}$ is
$H^2_{00}=-2$ and $H^2$ vanishes.

\subsection{$\Ac_3$ gauged axially by sp$\{T_0\}$}
(eq. (\ref{Sa3g0a}))

\noindent The coordinates of the target manifold are
\[ \{x_1,x_2,x_3\}; \]
the background fields are:
\beq
   G_{\mu\nu}=p\left(\begin{array}{ccc}
    0    & 2b   & -x_2 \\
    2b   & 0    & -x_1 \\
    -x_2 & -x_1 & -2
  \end{array}\right) \hsc
   B_{\mu\nu}=p\left(\begin{array}{ccc}
    0       & x_1x_2 & x_2  \\
    -x_1x_2 & 0      & -x_1 \\
    -x_2    & x_1    & 0
  \end{array}\right)
\eeq
\[ \Phi=-\half\log|x_1x_2+2b|+\mbox{const.} \]
where
\beq p=\frac{1}{x_1x_2+2b}; \eeq
the metric has determinant $\det G_{\mu\nu}=4bp^2$ and signature%
\footnote{The signature can be determined as follows. It can change
  only where the determinant vanishes or depends non-continuously on the
  coordinates. This happens only for $x_1x_2=-2b$, so it is enough to
  check the signature for $x_1=x_2=0$ and for $x_1x_2/b\goto\infty$.
  The first case is trivial. For the second case, we move to
  coordinates $(x,y,z)=(x_3,x_1x_2,x_1/x_2)$. The metric in these
  coordinates is
  \[ \frac{p}{y}\left(\begin{array}{ccc}
     -2y & -y & 0 \\
     -y  & b  & 0 \\
     0   & 0  & -b\left(\frac{y}{z}\right)^2
     \end{array}\right),
  \]
  so it has an eigenvalue with sign opposite to $b$. This rules out the
  signature sign$(b)(+++)$, which is the only other possibility
  compatible with the sign of the determinant.}
$(+,-{\rm sign}(b),-)$, so the signature of the gauged model is
obtained (when the gauging is non-singular) by ``subtracting'' the
signature of $dH$ from that of $dG$; all background fields are
singular at $x_1x_2=-2b$.

\noindent The inverse metric is
\beq
  G^{\mu\nu}=p\left(\begin{array}{ccc}
    -\frac{x_1^2}{4b}   & 1+\frac{x_1x_2}{4b} & -\half x_1 \\
    1+\frac{x_1x_2}{4b} & -\frac{x_2^2}{4b}   & -\half x_2 \\
    -\half x_1          & -\half x_2          & -b
  \end{array}\right);
\eeq
the Ricci tensor is
\beq
  R_{\mu\nu}=p^3\left(\begin{array}{ccc}
    0              & 8b^2           & x_2(x_1x_2-2b) \\
    8b^2           & 0              & x_1(x_1x_2-2b) \\
    x_2(x_1x_2-2b) & x_1(x_1x_2-2b) & 2(x_1x_2-2b)
  \end{array}\right)
\eeq
and the Ricci scalar is
\beq R=2p^2(5b-x_1x_2). \eeq
The torsion of the anti-symmetric tensor is $H_{123}=-4bp^2$;
\beq
  H^2_{\mu\nu}=p^3\left(\begin{array}{ccc}
    0      & 16b^2  & -8bx_2 \\
    16b^2  & 0      & -8bx_1 \\
    -8bx_2 & -8bx_1 & -16b
  \end{array}\right)
\eeq
and $H^2=24bp^2$.

\subsection{$\Ac_3$ gauged vectorially by sp$\{T_0\}$}
(eq. (\ref{Sa3g0v}))

\noindent The coordinates of the target manifold are
\[ \{x_0,x_2,x_3\}; \]
the background fields are
\beq
  G_{\mu\nu}=\frac{1}{x_2}\left(\begin{array}{ccc}
    -b(x_2+2b) & -b & -2b \\
    -b         & 0  & -1 \\
    -2b        & -1 & -2
  \end{array}\right) \hsc
   B_{\mu\nu}=\frac{1}{x_2}\left(\begin{array}{ccc}
    0     & b & x_2  \\
    -b    & 0 & -1    \\
    -x_2 & 1 & 0
  \end{array}\right)
\eeq
\[ \Phi=-\half\log|x_2|+\mbox{const.}; \]
the metric has determinant $\det G_{\mu\nu}=4b/x_2^2$ and signature%
\footnote{The other possibility ${\rm sign}(b)(+,+,+)$ is ruled out
  because, as in the axial case, the signature  does not depend on
  $|x_2|$ and the sum of the eigenvalues is tr$G_{\mu\nu}=-b+\Oc(1/x_2)$,
  which for $|x_2|$ large enough has a sign opposite to $b$.}
$(+,-{\rm sign}(b),-)$, as in the axial gauging;

\noindent the inverse metric is
\beq
   G^{\mu\nu}=\left(\begin{array}{ccc}
     -\frac{1}{b} & 0    & 1    \\
     0            & 2x_2 & -x_2 \\
     1            & -x_2 & -b
   \end{array}\right);
\eeq
the non-vanishing components of the Riemann tensor are (up to symmetry)
\beq
  R_{0202} = \frac{b^2}{x_2^3} \hsc
  R_{0232} = \frac{b}{x_2^3} \hsc
  R_{3232} = \frac{1}{x_2^3};
\eeq
the Ricci tensor is
\beq
   R_{\mu\nu}=\frac{1}{x_2^2}\left(\begin{array}{ccc}
     2b^2 & b & 2b \\
     b    & 0 & 1  \\
     2b   & 1 & 2
   \end{array}\right)
\eeq
and the Ricci scalar is $R=-\frac{2}{x_2}$.

\noindent The torsion of the anti-symmetric tensor vanishes.

\subsection{$\Ac_3$ gauged by null sp$\{T_0\}$}
(eq. (\ref{Sa3g0n}))

\noindent The coordinates of the target manifold are
\[ \{x_2,x_3\}; \]
the background fields are:
\beq
  G_{\mu\nu}=-\frac{1}{x_2}\left(\begin{array}{cc}
    0 & 1  \\
    1 & 2
  \end{array}\right) \hsc
  B_{\mu\nu}=\frac{1}{x_2}\left(\begin{array}{cc}
    0 & -1  \\
    1 &  0
  \end{array}\right),
\eeq
\[ \Phi=-\half\log|x_2|+\mbox{const.}; \]
the metric has signature $(+,-)$, which means that this gauging eliminated
one positive and one negative eigenvalue.

\noindent The inverse metric is
\beq
  G^{\mu\nu}=x_2\left(\begin{array}{cc}
    2  & -1 \\
    -1 & 0
  \end{array}\right);
\eeq
the Riemann tensor is $R_{2323}=\frac{1}{x_2^3}$;
the Ricci tensor is
\beq
  R_{\mu\nu}=\frac{1}{x_2^2}\left(\begin{array}{cc}
    0 & 1 \\
    1 & 2
  \end{array}\right)
\eeq
and the Ricci scalar is $R=-\frac{2}{x_2}$.

\noindent The torsion of the anti-symmetric tensor vanishes.

\subsection{$\Ac_6$ ungauged}
(eq. (\ref{Sa6}))

\noindent The coordinates of the target manifold are
\[ \{x_0,x_1,x_2,x_3,x_4,x_5,x_6\}; \]
the background fields are
\beq
  G_{\mu\nu}=\left(\begin{array}{ccccccc}
  b      & 0   & 0 & x_1x_2 & -x_2 & x_1 & 1 \\
  0      & 0   & 0 & x_2    & 0    &   1 &   \\
  0      & 0   & 0 & 0      & 1    &     &   \\
  x_1x_2 & x_2 & 0 & 1      &      &     &   \\
  -x_2   & 0   & 1 &        &      &     &   \\
  x_1    & 1   &   &        &      &     &   \\
  1      &     &   &        &      &     &
  \end{array}\right)
  \begin{picture}(0,0)\put(-20,-10){\Huge 0}\end{picture}.
\eeq
\beq
  B_{\mu\nu}=\left(\begin{array}{ccccccc}
  0      & 0 & 0    & -x_1x_2 & x_2 & -x_1 & 0 \\
  0      & 0 & 0    & 0       & 0   &  0   &   \\
  0      & 0 & 0    & x_1     & 0   &      &   \\
  x_1x_2 & 0 & -x_1 & 0       &     &      &   \\
  -x_2   & 0 & 0    &         &     &      &   \\
  x_1    & 0 &      &         &     &      &   \\
  0      &   &      &         &     &      &
  \end{array}\right)
  \begin{picture}(0,0)\put(-20,-10){\Huge 0}\end{picture},
\eeq
\beq \Phi=0;\eeq
the inverse metric is
\beq
  G^{\mu\nu}=\left(\begin{array}{ccccccc}
      &      &     &      &   &       & 1    \\
      &      &     &      &   & 1     & -x_1 \\
      &      &     &      & 1 & 0     & x_2  \\
      &      &     & 1    & 0 & -x_2  & 0    \\
      &      & 1   & 0    & 0 & 0     & 0    \\
      & 1    & 0   & -x_2 & 0 & x_2^2 & 0    \\
    1 & -x_1 & x_2 & 0    & 0 & 0     & -b
  \end{array}\right)
  \begin{picture}(0,0)\put(-50,5){\Huge 0}\end{picture},
\eeq
the non-vanishing components of the Riemann tensor are (up to symmetry)
\[ R_{0103}=-\frac{x_2}{4} \hsc R_{0202}=\frac{x_1^2}{4}, \]
\beq R_{0203}=R_{0212}=\frac{x_1}{4}, \eeq
\[ R_{0150}=R_{0123}=R_{0240}=R_{0213}=R_{1212}=\frac{1}{4}; \]
the only non-vanishing component of the Ricci tensor is $R_{00}=-1$
and the Ricci scalar vanishes.

\noindent The non-vanishing components of the torsion are (up to symmetry)
\beq H_{013}=x_2 \hsc H_{023}=x_1, \eeq
\[ H_{015}=H_{042}=H_{123}=1 \]
and it is covariantly constant;
\noindent the only non-vanishing component of $H^2_{\mu\nu}$ is
$H^2_{00}=-4$ and $H^2$ vanishes.

\subsection{$\Ac_6$ gauged by sp$\{T_6\}$}
(eq. (\ref{Sa6g6}))

\noindent The coordinates of the target manifold are
\[ \{x_1,x_2,x_3,x_4,x_5\}; \]
the background fields are
\beq
  G_{\mu\nu}=\left(\begin{array}{ccccc}
  0   & 0 & x_2 & 0 & 1 \\
  0   & 0 & 0   & 1 &   \\
  x_2 & 0 & 1   &   &   \\
  0   & 1 &     &   &   \\
  1   &   &     &   &
  \end{array}\right)
  \begin{picture}(0,0)\put(-14,-8){\Huge 0}\end{picture},
\eeq
\beq
  B_{\mu\nu}=\left(\begin{array}{ccccc}
  0 & 0    & 0   & 0 & 0 \\
  0 & 0    & x_1 & 0 &   \\
  0 & -x_1 & 0   &   &   \\
  0 & 0    &     &   &   \\
  0 &      &     &   &
  \end{array}\right)
  \begin{picture}(0,0)\put(-14,-8){\Huge 0}\end{picture},
\eeq
\beq \Phi=0;\eeq
the inverse metric is
\beq
  G^{\mu\nu}=\left(\begin{array}{ccccc}
      &   &      &   & 1     \\
      &   &      & 1 & 0     \\
      &   & 1    & 0 & -x_2  \\
      & 1 & 0    & 0 & 0     \\
    1 & 0 & -x_2 & 0 & x_2^2 \\
  \end{array}\right)
  \begin{picture}(0,0)\put(-35,3){\Huge 0}\end{picture}.
\eeq
the only non-vanishing component of the Riemann tensor is (up to symmetry)
$R_{1212}=\frac{1}{4}$, and the Ricci tensor vanishes.

\noindent The torsion of the anti-symmetric tensor is
\beq H=6dx_1\wedge x_2\wedge x_3 \eeq
(which means $H_{123}=1$) and it is covariantly constant.

\noindent $H^2_{\mu\nu}$ vanishes.

\subsection{$\Ac_6$ gauged by sp$\{T_6,T_m\}$}
(eq. (\ref{Sa6g642}))

\noindent The coordinates of the target manifold are
\[ \{x_1,x_3,x_4,x_5\}; \]
The background fields are:
\beq
  G_{\mu\nu}=\left(\begin{array}{cccc}
    0            & 2px_1x_3+x_2 & -2px_3 & 1 \\
    2px_1x_3+x_2 & 1            & -2px_1 & 0 \\
    -2px_3       & -2px_1       & 4p     & 0 \\
    1            & 0            & 0      & 0
  \end{array}\right),
\eeq
\beq
   B_{\mu\nu}=2p\left(\begin{array}{cccc}
  0      & -x_1x_3 & x_3  & 0 \\
  x_1x_3 & 0       & -x_1 & 0 \\
  -x_3   & x_1     & 0    & 0 \\
  0      & 0       & 0    & 0
  \end{array}\right),
\eeq
\beq \Phi=-\half\log(x_1^2+a)+\mbox{const.} \eeq
where
\[ p=\frac{1}{x_1^2+a}; \]
the metric has determinant $\det G_{\mu\nu}=-4ap^2$ and signature
$(+,+,{\rm sign}(a),-)$, so again the signature of the gauged model is
obtained (when the gauging is non-singular) by ``subtracting'' the
signature of $dH$ from that of $dG$.

\noindent The inverse metric is
\beq
  G^{\mu\nu}=\frac{1}{a}\left(\begin{array}{cccc}
    0  & 0               & 0                        & a                  \\
    0  & q_+             & \half q_+x_1             & -(x_1x_3+q_+x_2)   \\
    0  & \half q_+x_1    & \frac{1}{4}q_+^2         &
                                                -\half(q_-x_3+q_+x_1x_2) \\
    a & -(x_1x_3+q_+x_2) & -\half(q_-x_3+q_+x_1x_2) &
                                                (x_3^2+2x_1x_2x_3+q_+x_2^2)
  \end{array}\right)
\eeq
with $q_\pm=x_1^2\pm a$;

\noindent the non-vanishing components of the Riemann tensor are (up to
symmetry)
\begin{eqnarray}
  \nonumber R_{1313} & = & -p^2(2x_1^2-3a)   \\
            R_{1314} & = & 2p^3x_1(2x_1^2-a) \\
  \nonumber R_{1414} & = & -4p^3(2x_1^2-a);
\end{eqnarray}
the only non-vanishing component of the Ricci tensor is
\beq R_{11}=-2p^2(x_1^2-2a) \eeq
and the Ricci scalar vanishes.
\noindent The torsion of the anti-symmetric tensor is
\beq H=-24ap^2dx^1\wedge dx^3\wedge dx^4 \eeq
which means $H_{134}=-4ap^2$; The only non-vanishing component of
$H^2_{\mu\nu}$ is $H^2_{11}=8ap^2$ and $H^2$ vanishes.

\subsection{$\Ac_6$ gauged by sp$\{T_6,T_2\}$}
(eq. (\ref{Sa6g62}))

\noindent The coordinates of the target manifold are
\[ \{x_1,x_4,x_5\}; \]
the background fields are
\beq
  G_{\mu\nu}=\left(\begin{array}{ccc}
    0 & 0               & 1 \\
    0 & \frac{4}{x_1^2} & 0 \\
    1 & 0               & 0
  \end{array}\right),
\eeq
\[ B_{\mu\nu}=0 \hsc \Phi=-\log(x_1)+\mbox{const.;} \]
the inverse metric is
\beq
  G^{\mu\nu}=\left(\begin{array}{ccc}
    0 & 0               & 1 \\
    0 & \frac{x_1^2}{4} & 0 \\
    1 & 0               & 0
  \end{array}\right);
\eeq
the only non-vanishing component of the Riemann tensor is (up to
symmetry) $R_{1441}=\frac{8}{x_1^4}$, the only non-vanishing component
of the Ricci tensor is $R_{11}=-\frac{2}{x_1^2}$ and the Ricci scalar
vanishes.





\begin{thebibliography}{99}


 \bibitem{WZNW}
%

   E.\ Witten,
   {\em Non-Abelian bosonisation in two dimensions},
   Commun.\ Math.\ Phys.\ {\bf 92} (1984) 455.

%

%

 \bibitem{GWZNW}



   W.\ Nahm, {\em Gauging symmetries of two-dimensional conformally
   invariant models}, Davis preprint UCD-88-02 (1988); 

   K.\ Bardakci, E.\ Rabinovici and B.\ S\"{a}ring,
   {\em String models with $c<1$ components},
   Nucl.\ Phys.\ {\bf B299} (1988) 151;


   K.\ Gaw\c{e}dzki and A.\ Kupiainen,
   {\em Coset construction from functional integrals}
   Nucl.\ Phys.\ {\bf B320} (1989) 625;


   D.\ Karabali, Q-H.\ Park, H.J.\ Schnitzer and Z.\ Yang,
   {\em A GKO construction based on a path integral formulation of
   gauged Wess-Zumino-Witten actions},
   Phys.\ Lett.\ {\bf B216} (1989) 307;





%














 \bibitem{Nappi-Witten} 
   C.R.\ Nappi and  E.\ Witten,
   {\em A Wess-Zumino-Witten model based on a non-semi-simple group},
   hep-th/9310112, Phys.\ Rev.\ Lett.\ {\bf 71} (1993) 3751.

 \bibitem{Kiritsis-Kounnas} 
   E.\ Kiritsis and C.\ Kounnas,
   {\em String propagation in gravitational wave backgrounds},
   hep-th-9310202, Phys.\ Lett.\ {\bf B320} (1994) 264. 

 \bibitem{ORS} 
   D.I.\ Olive, E.\ Rabinovici and A.\ Schwimmer,
   {\em A class of string backgrounds as a semiclassical limit of WZW
   models}, hep-th/9311081, Phys.\ Lett.\ {\bf B321} (1994) 361. 

 \bibitem{Sfetsos} 
   K.\ Sfetsos,
   {\em Gauging a non-semi-simple WZW model},
   hep-th/9311010, Phys.\ Lett.\ {\bf B324} (1994) 335. 
   {\em Exact string backgrounds from WZW models based on
   non-semi-simple groups},
   hep-th/9311093, Int.\ J.\ Mod.\ Phys.\ {\bf A9} (1994) 4759; 

 \bibitem{Sfet-gauge} 
   {\em Gauged WZW models and non-Abelian duality},
   hep-th/9402031, Phys.\ Rev.\ {\bf D50} (1994) 2784.

 \bibitem{Kehagias-Meessen} 
   A.A\ Kehagias and P.A.A.\ Meessen, {\em Exact string background
   from a WZW model based on the Heisenberg group},
   hep-th/9403041, Phys.\ Lett.\ {\bf B331} (1994) 77.

 \bibitem{Antoniadis-Obers} 
   I.\ Antoniadis and N.\ Obers,
   {\em Plane gravitational waves in string theory},
   hep-th/9403191, Nucl.\ Phys.\ {\bf B423} (1994) 639.

 \bibitem{Kiritsis-Kounnas-Lust} 
   E.\ Kiritsis, C.\ Kounnas and D.\ L\"{u}st,
   {\em Superstring gravitational wave backgrounds with spacetime
   supersymmetry}, hep-th/9404114, Phys.\ Lett.\ {\bf B331} (1994) 321.

 \bibitem{Mohammedi-0-1} 
   N.\ Mohammedi, {\em On bosonic and supersymmetric current algebras
   for non-semi-simple Lie groups},
   hep-th/9312182, Phys.\ Lett.\ {\bf B325} (1994) 371.

 \bibitem{Kehagias}
   A.A.\ Kehagias, {\em All WZW models in $D\le5$},
   hep-th/9406136. 





 \bibitem{Tseytlin-9505} 
   A.A.\ Tseytlin, {\em On gauge theories for non-semi-simple groups},
   hep-th/9505129. 

 \bibitem{FOF}
   J.M.\ Figueroa-O'Farrill and S.\ Stanciu,
   {\em Non-reductive WZW models and their CFTs}, hep-th/9506151;
   {\em On the structure of symmetric self-dual Lie algebras},
   hep-th/9506152.


 \bibitem{FOF-Stan} 
   J.M.\ Figueroa-O'Farrill and S.\ Stanciu
   {\em Non-semi-simple Sugawara constructions},
   hep-th/9402035, Phys.\ Lett.\ {\bf B327} (1994) 40. 





 \bibitem{Ino-Wig}
   E.\ Inonu and E.P.\ Wigner,
   {\em On the contraction of groups and their representations},
   Proc.\ Natl.\ Acad.\ Sci.\ U.S.\ {\bf 39} (1953) 510.

 \bibitem{Med-Rev}
   A.\ Medina and Ph.\ Revoy,
   {\em Alg\`{e}bres de Lie et produit scalaire invariant},
   Ann.\ Scient.\ \'{E}c.\ Norm.\ Sup.\ {\bf 18} (1985) 553.

 \bibitem{OSP}
   O.\ Pelc, in preparation.

 \bibitem{Null}
   C.\ Klim\v{c}\'{\i}k and A.A.\ Tseytlin,
   {\em Exact four-dimensional string solutions and Toda-like sigma
   models from `null-gauged' WZNW theories},
   hep-th/9402120, Nucl.\ Phys.\ {\bf B424} (1994) 71.

 \bibitem{beta}
   C.G.\ Callan, D.\ Friedan, E.J.\ Martinec and M.J.\ Perry,
   {\em Strings in background fields},
   Nucl.\ Phys.\ {\bf B262} (1985) 593.





 \bibitem{GWZNW-NLSM}
   K.\ Bardakci, M.\ Crescimanno and E.\ Rabinovici,
   {\em Parafermions from coset models},
   Nucl.\ Phys.\ {\bf B344} (1990) 344;

   I.\ Bars and D.\ Nemeschansky,
   {\em String propagation in backgrounds with curved space-time},
   Nucl.\ Phys.\ {\bf B348} (1991) 89;

   E.\ Witten, {\em String theory and black holes},
   Phys.\ Rev.\ {\bf D44} (1991) 314.

%
%
%
%



 \bibitem{Dilaton}
   T.H.\ Buscher,
   {\em A symmetry of the string background field equations},
   Phys.\ Lett.\ {\bf B194} (1987) 59; 
   {\em Path-integral derivation of quantum duality in non-linear
   sigma-models}, Phys.\ Lett.\ {\bf B201} (1988) 466.

 %
 %

 \bibitem{Mult-Anomaly}
   A.\ Giveon, and M.\ Ro\u{c}ek, {\em On non-Abelian duality},
   hep-th/9308154, Nucl.\ Phys.\ {\bf B421} (1994) 173; 

   E.\ Alvarez, L.\ Alvarez-Gaum\'{e} and Y.\ Lozano,
   {\em On non-Abelian duality},
   hep-th/9403155, Nucl.\ Phys.\ {\bf B424} (1994) 155; 

   S.\ Elitzur, A.\ Giveon, E.\ Rabinovici, A.\ Schwimmer and
   G.\ Veneziano, {\em Remarks on non-Abelian duality},
   hep-th/9409011, Nucl.\ Phys.\ {\bf B435} (1995) 147.

 \bibitem{Duality}
   For a review, see:
   A.\ Giveon, M.\ Porrati, and E.\ Rabinovici,
   {\em Target space duality in string theory},
   hep-th/9401139, Phys.\ Rep.\ {\bf 244} (1994) 77. 

 \bibitem{Hau-Sch}
   M.\ Hausner and J.T.\ Schwartz,
   {\em Lie groups and their Lie algebras}, Gordon and Breach 1968.

 \bibitem{Exact-compare} 
   C.\ Klim\v{c}\'{\i}k and A.A.\ Tseytlin,
   {\em Duality invariant class of exact string backgrounds},
   hep-th/9311012, Phys.\ Lett.\ {\bf B323} (1994) 305;

   K.\ Sfetsos and A.A.\ Tseytlin, {\em Four-dimensional plane wave
   string solutions with coset CFT description},
   hep-th/9404063, Nucl.\ Phys.\ {\bf B427} (1994) 245.

%

%
%

\end{thebibliography}
\end{document}